\documentclass[pdflatex,sn-mathphys-num]{sn-jnl}

\usepackage{graphicx}%
\usepackage{multirow}%
\usepackage{multicol}%
\usepackage{amsmath,amssymb,amsfonts}%
\usepackage{amsthm}%
\usepackage{mathrsfs}%
\usepackage[title]{appendix}%
\usepackage{xcolor}%
\usepackage{textcomp}%
\usepackage{manyfoot}%
\usepackage{booktabs}%
\usepackage{algorithm}%
\usepackage{algorithmicx}%
\usepackage{algpseudocode}%
\usepackage{listings}%
\usepackage{cleveref}
\usepackage{siunitx}
\usepackage{subcaption}
\usepackage{float}
\usepackage{geometry}%
\newgeometry{margin=1in}
\crefname{algocf}{alg.}{algs.}
\Crefname{algocf}{Algorithm}{Algorithms}

\theoremstyle{thmstyleone}%
\theoremstyle{thmstyletwo}%

\theoremstyle{thmstylethree}%

\begin{document}

\title[Response Estimation and System Identification of Dynamical Systems via Physics-Informed Neural Networks]{Response Estimation and System Identification of Dynamical Systems via Physics-Informed Neural Networks}


\author*[1]{\fnm{Marcus} \sur{Haywood-Alexander}}\email{mhaywood@ethz.ch}

\author[1]{\fnm{Giacomo} \sur{Arcieri}}\email{garcieri@ethz.ch}

\author[1]{\fnm{Antonios} \sur{Kamariotis}}\email{akamariotis@ethz.ch}

\author[1,2]{\fnm{Eleni} \sur{Chatzi}}\email{echatzi@ethz.ch}

\affil*[1]{\orgdiv{Department of Civil, Environmental and Geomatic Engineering}, \orgname{ETH Zürich}, \orgaddress{\street{Wolfgang-Pauli Strasse}, \city{Zürich}, \postcode{8049}, \country{Switzerland}}}

\affil[2]{\orgdiv{Future Resilient Systems}, \orgname{Singapore-ETH Centre}, \orgaddress{\country{Singapore}}}

\abstract{The accurate modelling of structural dynamics is crucial across numerous engineering applications, such as Structural Health Monitoring (SHM), seismic analysis, and vibration control. Often, these models originate from physics-based principles and can be derived from corresponding governing equations, often of differential equation form. However, complex system characteristics, such as nonlinearities and energy dissipation mechanisms, often imply that such models are approximative and often imprecise. This challenge is further compounded in SHM, where sensor data is often sparse, making it difficult to fully observe the system's states. To address these issues, this paper explores the use of Physics-Informed Neural Networks (PINNs), a class of physics-enhanced machine learning (PEML) techniques, for the identification and estimation of dynamical systems. PINNs offer a unique advantage by embedding known physical laws directly into the neural network's loss function, allowing for simple embedding of complex phenomena, even in the presence of uncertainties. This study specifically investigates three key applications of PINNs: state estimation in systems with sparse sensing, joint state-parameter estimation, when both system response and parameters are unknown, and parameter estimation within a Bayesian framework to quantify uncertainties. The results demonstrate that PINNs deliver an efficient tool across all aforementioned tasks, even in presence of modelling errors. However, these errors tend to have a more significant impact on parameter estimation, as the optimization process must reconcile discrepancies between the prescribed model and the true system behavior. Despite these challenges, PINNs show promise in dynamical system modeling, offering a robust approach to handling uncertainties. }




\maketitle

\section{Introduction}\label{sec:intro}

The characteristics of structural system dynamics are often crucial for inference of features that serve for system diagnostic and prognostic tasks within the context of structural health monitoring (\emph{SHM}) campaigns \cite{Farrar-Worden07,dervilis2019nonlinear}. %
Beyond \emph{SHM}, structural identification of dynamical systems plays a vital role in seismic analysis \cite{reuland2023comparative}, design and optimisation \cite{felkner2013interactive}, vibration control \cite{chondrogiannis2023design}, and aeroelasticity \cite{abdallah2017fatigue}. %
Besides feature detection, a typical downstream task is estimating the response of vibrating systems \cite{vettori2023adaptive}, which is valuable for structural safety and comfort \cite{bao2016case}, noise control \cite{mao2013control}, and dynamic load mitigation \cite{manimala2014dynamic}. %
Moreover, response prediction can compliment or serve as an alternative to feature identification for the aforementioned motivations. %

Traditionally, schemes addressing these motivations rely on physics-based models, including hybrid methods that combine first principles with data \cite{haywood2023discussing}. %
For such methods to be effective, it is essential to adopt accurate and robust models in order to reduce (or eliminate) detection or prediction errors . %
State, parameter and input estimations can be succeeded via Bayesian filtering schemes, either online or offline\cite{tatsis2022sequential,azam2015dual,champneys2024bindy}, which, given a sufficiently accurate system model, can provide reliable and reasonably fast estimations.
Modern engineering systems often involve complex materials, geometries and intricate energy harvesting and vibration mitigation mechanisms, which may be associated with complex mechanics and failure patterns \cite{duenas2009cascading,van2012computational,kim2017failure}. %
This results in behaviour that cannot be easily described purely by data observations or common simplified modelling assumptions, and these forward models, often grounded in engineering knowledge, may suffer from inaccuracies due to uncertainties and incomplete information about system characteristics \cite{erazo2015bayesian}. %

In many applied engineered systems, the level of hardware (sensors) required for full state observation is often not feasible, and the hardware redundancy required for data-driven approaches is often not practical. %
Therefore, when employing modern smart systems, it is useful to be able to extrapolate the domain of prediction beyond that of which measurements are available, particularly in the spatial domain \cite{papatheou2023virtual}. %
This is particularly useful in the context of SHM, where the number of sensors is often limited due to cost, installation complexity, and fault-susceptibility \cite{ercan2022optimal,vettori2020virtual}. %
For parameter estimation or system identification, a deterministic estimate is often viable for us in downstream tasks, however, a more complete description requires the attachment of the uncertainty involved in the delivered estimate \cite{Farrar-Worden07,kamariotis2023off}. %

Recent advances in computing power and data availability have led to the increased use of purely data-driven (black-box) machine learning tools to address model form uncertainties in complex engineering systems \cite{reich1997machine,cuomo2022scientific}. %
However, these methods often lack generalization, being effective primarily within the specific domains from which the training data were collected, as well as, in their simplest form, offering a ``best'' deterministic estimate of the solution. %
In modelling complex systems, there is a need for a balanced approach that combines physics-based and data-driven models \cite{pawar2021model}, to improve capture of complex mechanisms, and to provide a mean estimate to allow uncertainty quantification. %
An effective approach to overcome the aforementioned pitfalls in modelling complex engineering systems is to integrate the physics-based aspect of forward modelling with the data-driven aspect of machine learning to account for modelling uncertainties and imprecision. %
By integrating known physical principles with machine learning, enhanced approaches can be developed that leverage prior physics knowledge to improve data-driven models. %
This fusion, referred to as \emph{physics-enhanced machine learning} (\emph{PEML}) \cite{faroughi2022physics,haywood2023discussing}, embeds prior physics knowledge into the learning processing, typically resulting in more interpretable models \cite{o2019physics,choudhary2020physics,xiaowei2021physics}. %

A notable example is the \emph{physics-informed neural network} (\emph{PINN}) \cite{raissi2019physics,moradi2023novel}, which has gained attention for its versatility and the ease with which complex models can be prescribed \cite{faroughi2022physics}. %
Additionally, this approach is reasonably flexible in application and can be adapted for various schemes \cite{stiasny2023physics}, such as model solution recovery from sparse data \cite{zhong2020symplectic}, parameter estimation \cite{zhang2021structural,parziale2024physics,maes2021observability}, and forward modeling \cite{haghighat2021deep}. %
In fact \emph{PINN}s have been shown to simultaneously estimate the state and parameters of dynamical systems \cite{zhang2024dual,moradi2023novel}, and face challenges when convergence rates of loss functions differ \cite{zhang2024physics}. %
\emph{PINN}s embed prior physics knowledge, in the form of partial or ordinary differential equations, by creating an additional loss (objective) function to be minimised. %
Similar to Bayesian filtering (BF) schemes, the results depend on the adequacy of the prescribed model, however, the schemes do also have distinct differences, as will be discussed in the next paragraph. %

One distinct difference between Bayesian filtering schemes and \emph{PINN}s is the mode of prediction; the former scheme can, and often is, adopted in an online prediction mode, where time history data is continuously used to update the predictor until convergence. %
Whereas \emph{PINN}s operate in a more limited manner in that they are formed as `instance' modellers; meaning they replicate the time history of a given scenario of initial conditions, system input, and boundary conditions. %
This means that \emph{PINN}s must operate in `batch mode' and thus would not be suitable for real-time estimation tasks, and instead for prediction in scenarios where there is a lack of confidence in the prescribed model. %
In the context of \emph{PINN}s for structural response estimation, they are employed without \emph{specifically} prescribing the initial conditions, thus, their reliance on them is less explicit, but still palpable. %
This is because the data-driven aspect of training implicitly relies on the initial conditions, and in training using physical laws, the \emph{PINN} prediction still relies on the trajectory of the system, defined, in part, by its initial state. %
Another difference is that as \emph{PINN}s operate using a weak-form objective function, and so uncertainty is captured by allowing a `best-estimate' to a deterministic solution, whereas in Bayesian filtering, the uncertainty is specifically captured with a stochastic representation. %
The result of the weak-form objective with uncertainties and model inaccuracies, as well as \emph{PINN} performance when the prescribed model deviates from the actual system behaviour, is not well-explored. %

The scope of this study encompasses three critical aspects of dynamical system modelling using \emph{PINN}s. %
First, \textbf{state estimation in a sparse sensing context} is explored, where the challenge lies in estimating unmeasured degrees of freedom (DOFs) within an extrapolated domain of the system. %
State estimation aims to reconstruct the system's states based on partial measurements/observations, and domain extrapolation is crucial for predicting the system's behavior beyond the observed data. %
This study aims to explore the accuracy and reliability of state estimation techniques leveraging \emph{PINN}s. %
Second, \textbf{state-parameter estimation} is addressed, a task where a subset of both system response and parameters are unknown. %
The objective is to apply \emph{PINN}s to this dual-faceted aim, integrating data and physical laws to refine model predictions. %
Furthermore, in the state and state-parameter estimation schemes, the influence of model accuracy are explored, by embedding model errors in the form of omitted nonlinearities and incorrect system parameter values. %
By doing so, the effect of the weak-form embedding of prior physics knowledge, in the context of estimating vibrating systems, is explored. %
Finally, exploitation of \emph{PINN} architectures for \textbf{parameter estimation in a Bayesian context} is performed, focusing on the role of parameter estimation in dynamic system modelling and employing Bayesian methods to quantify uncertainties. %
The goal is to integrate \emph{PINN}s with Bayesian approaches to enhance parameter estimation, providing a probabilistic framework to capture uncertainty, going beyond classic deterministic estimates available with \emph{PINN}s. %

The remainder of the paper begins with a description of the multi-degree-of-freedom (MDOF) system and its state-space representation used throughout the study and details of the simulated data. %
This is followed by a general description of the \emph{PINN} approach and how it can be coupled to the MDOF state-space system, including its specific application in the context of the three schemes described above. %
Next, results of these schemes applied to the simulated model are presented and discussed, along with a comparison of the effects of model errors. %
The paper concludes with a summary of the promises, challenges, and future research directions for the application of \emph{PINN}s in the identification and estimation of vibrating structures. %

\section{Multi-Degree of Freedom System Model}\label{sec:mdof-system}

\begin{figure}[h!]
    \centering
    \includegraphics[width=0.9\textwidth]{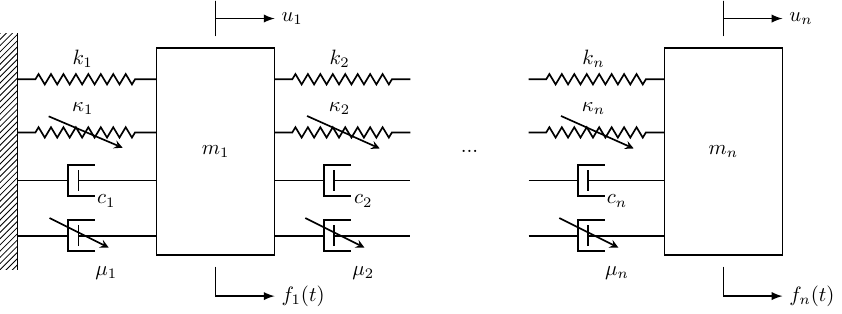}
    \caption{MDOF Nonlinear Oscillator}
    \label{fig:ndof_nonlin_diag}
\end{figure}

In this work, a generic multi-degree of freedom (MDOF) oscillator is employed as a working example. %
This is defined as sequentially connected lumped-mass system of $n$ degrees of freedom, fixed at the first (left) connection, as shown in \Cref{fig:ndof_nonlin_diag}. %
The nonlinearities are introduced in the form of a Duffing spring with constant $\kappa$, and Van Der Pol damping with constant $\mu$, which have respective restoring forces $r$ of,
\begin{subequations}
    \begin{equation}
        r_{\kappa} = \kappa x^3
    \end{equation}
    \begin{equation}
        r_{\mu} = \mu(x^2 - 1) \dot{x}
    \end{equation}
    \label{eqs:nonlin_forces}
\end{subequations}
\noindent where $x$ and $\dot{x}$ are the relative displacement and velocity of the spring/damper, respectively. The system model is then defined by,
\begin{equation}
    \mathbf{M}(\boldsymbol{\theta_s})\ddot{\mathbf{u}}(t) + \mathbf{C}(\boldsymbol{\theta_s})\dot{\mathbf{u}}(t) + \mathbf{C}_n(\boldsymbol{\theta_s}) g_c({\mathbf{u}(t),\dot{\mathbf{u}}}(t)) + \mathbf{K}(\boldsymbol{\theta_s})\mathbf{u}(t) + \mathbf{K}_n(\boldsymbol{\theta_s}) g_k(\mathbf{u}(t)) = \mathbf{f}(t)
    \label{eq:governing_eq_nom}
\end{equation}
\noindent where $(\boldsymbol{\theta_s})$ is the vector of system parameters, $\mathbf{K}$ and $\mathbf{K}_n$ are the linear and nonlinear stiffness matrices, $\mathbf{C}$ and $\mathbf{C}_n$ are the linear and nonlinear damping matrices, and $\mathbf{M}$ is the mass matrix. %
The stiffness, damping, nonlinear stiffness, nonlinear damping, and mass matrices are trivial to determine, and, for brevity, are given for this system in the appendix. %
From \Cref{eqs:nonlin_forces}, the nonlinear functions of the state are,
\begin{equation}
    g_c({\mathbf{u},\dot{\mathbf{u}}}) = \begin{bmatrix}
        (u_1^2-1)\dot{u}_1 \\
        ((u_2 - u_1)^2-1)(\dot{u}_2 - \dot{u}_1) \\
        \vdots \\
        ((u_n - u_{n-1})^2-1)(\dot{u}_n - \dot{u}_{n-1}) \\
    \end{bmatrix}, \quad
    g_k({\mathbf{u}}) = \begin{bmatrix}
        \dot{u}_1^3 \\
        (\dot{u}_2 - \dot{u}_1)^3 \\
        \vdots \\
        (\dot{u}_n - \dot{u}_{n-1})^3 \\
    \end{bmatrix}
\end{equation}
The state space representation of \Cref{eq:governing_eq_nom} is,
\begin{subequations}
    \begin{equation}
        \dot{\mathbf{z}}(t) = \mathbf{A}(\boldsymbol{\theta_s})\mathbf{z}(t) + \mathbf{A}_n(\boldsymbol{\theta_s})\mathbf{g}_n(t) + \mathbf{H}(\boldsymbol{\theta_s})\mathbf{f}(t)
        \label{eq:governing_eq_state}
    \end{equation}
    \begin{equation}
        \mathbf{y}(t) = \mathbf{B}(\boldsymbol{\theta_s})\mathbf{z}(t) + \mathbf{B}_n(\boldsymbol{\theta_s})\mathbf{g}_n(t) + \mathbf{D}(\boldsymbol{\theta_s})\mathbf{f}(t)
        \label{eq:state_eq_output}
    \end{equation}
\end{subequations}
\noindent where $\mathbf{z}$ is the state variable, $\mathbf{g}_n$ is the nonlinear function of the state, $\mathbf{f}$ is the system input (force), and $\mathbf{y}$ is the vector of (sparse) observations,
\begin{equation}
    \mathbf{z} = \left\{ \mathbf{u}^T, \dot{\mathbf{u}}^T \right\}^T, \quad
    \mathbf{g}_n = \left\{ g_k(\mathbf{u})^T,\; _c(\mathbf{u},\dot{\mathbf{u}})^T \right\}^T, \quad
    \mathbf{f} = \left\{ f_1, f_2, ... , f_n \right\}^T, \quad
    \mathbf{y} = \left\{ \mathbf{u}^T, \dot{\mathbf{u}}^T, \ddot{\mathbf{u}}^T \right\}^T
\end{equation}
The matrices of the state equation are,
\begin{equation}
    \begin{split}
    \mathbf{A}(\boldsymbol{\theta_s}) &= \begin{bmatrix} \boldsymbol{0} & \mathbf{I} \\ -\mathbf{M}^{-1}(\boldsymbol{\theta_s})\mathbf{K}(\boldsymbol{\theta_s}) & -\mathbf{M}^{-1}(\boldsymbol{\theta_s})\mathbf{C}(\boldsymbol{\theta_s}) \end{bmatrix}, \\
    \mathbf{A}_n(\boldsymbol{\theta_s}) &= \begin{bmatrix} \boldsymbol{0} & \boldsymbol{0} \\ -\mathbf{M}^{-1}(\boldsymbol{\theta_s})\mathbf{K}_n(\boldsymbol{\theta_s}) & -\mathbf{M}^{-1}(\boldsymbol{\theta_s})\mathbf{C}_n(\boldsymbol{\theta_s}) \end{bmatrix} \\
    \mathbf{H}(\boldsymbol{\theta_s}) &= \begin{bmatrix} \boldsymbol{0} \\ \mathbf{M}^{-1}(\boldsymbol{\theta_s}) \end{bmatrix}
    \end{split}
\end{equation}
and the matrices of the output equation are,
\begin{equation}
    \begin{split}
    \mathbf{B} &= \begin{bmatrix} \mathbf{S}_d & \boldsymbol{0} \\ \boldsymbol{0} & \mathbf{S}_v \\ -\mathbf{S}_a\mathbf{M}^{-1}(\boldsymbol{\theta_s})\mathbf{K}(\boldsymbol{\theta_s}) & -\mathbf{S}_a\mathbf{M}^{-1}(\boldsymbol{\theta_s})\mathbf{C}(\boldsymbol{\theta_s}) \end{bmatrix}, \\
    \mathbf{B}_n &= \begin{bmatrix} \boldsymbol{0} & \boldsymbol{0} \\ \boldsymbol{0} & \boldsymbol{0} \\ -\mathbf{S}_a\mathbf{M}^{-1}(\boldsymbol{\theta_s})\mathbf{K}_n(\boldsymbol{\theta_s}) & -\mathbf{S}_a\mathbf{M}^{-1}(\boldsymbol{\theta_s})\mathbf{C}_n(\boldsymbol{\theta_s}) \end{bmatrix}, \\
    \mathbf{D} &= \begin{bmatrix} \boldsymbol{0} \\ \boldsymbol{0} \\ \mathbf{S}_a\mathbf{M}^{-1}(\boldsymbol{\theta_s}) \end{bmatrix}
    \end{split}
\end{equation}
The matrices $\mathbf{S}_d$, $\mathbf{S}_v$, and $\mathbf{S}_a$ are the selection matrices for displacement, velocity, and acceleration, respectively, which define which DOFs are measured for each variable. %
These matrices deliver the measurement vector $\mathbf{y}$, through the output \Cref{eq:state_eq_output}. %

\subsection{Simulated MDOF Model}

For the running examples employed in this paper, several MDOF systems were simulated, with the number of degrees of freedom varying in an effort to explore the effect of dimensionality, with the forcing $\mathbf{f}$ generated as white Gaussian noise, banded between frequencies of 0.05 and 1.0 Hz, with an amplitude of 1.0 N, applied to the first degree of freedom, i.e.\ $f_{1} (t) = \textrm{GWN}(0.5\textrm{rad}, 4\textrm{rad}, 60\textrm{s})$. %
For all systems the values of the parameters are kept constant at $m_i=10.0$ \si{\kilogram}, $c_i=1.0$ \si{\newton\second\per\metre}, and $k_i=15.0$ \si{\newton\per\metre}, and the nonlinear components, when included, are set at $\kappa_i=100.0$ \si{\newton\per\metre^3}, and $\mu_i=0.75 \si{\newton\second\per\metre^3}$. %
The systems were then simulated with a fourth-order Runge-Kutta integration \cite{butcher2016numerical} using \Cref{eq:governing_eq_state}, over a 60s window with 4096 samples and Gaussian noise added to the signal controlled by a representative signal-to-noise ratio (SNR). %
These noisy data are then used as the ground truth for the examples shown throughout the paper. %


\section{Physics-Informed Neural Networks}\label{sec:pinns}

Most modern ML methods are based on use or extension of the neural network (NN), which can be used as a universal function approximator. %
For a regression problem, the aim of an NN is to determine an estimate of the mapping from the input $\mathbf{x}$, to the output $\mathbf{y}$, i.e.\ $\tilde{\mathbf{y}} = \mathcal{N}_{\mathbf{y}}(\mathbf{x}; \mathbf{W}, \mathbf{B})$, where $\mathbf{W}$ and $\mathbf{B}$ are the weights and biases of the network, respectively. %
The aim of the training stage is to then determine the network parameters $\boldsymbol{\theta}_n = \{\mathbf{W},\mathbf{B}\}$, by minimising an objective function defined so that when the value vanishes, the solution is satisfied. %
\begin{equation}
    L_o(\mathbf{x}; \boldsymbol{\theta}_n) = \left\langle \mathbf{y}^* - \mathcal{N}_{\mathbf{y}}(\mathbf{x}; \boldsymbol{\theta}_n) \right\rangle _{\Omega_o}, \qquad
    \langle \bullet \rangle _{\Omega_{\kappa}} = \frac{1}{N_{\kappa}}\sum_{x\in\Omega_{\kappa}}||\bullet||^2
    \label{eq:obs_loss}
\end{equation}

If the physics of the system is known (or estimated) in the form of ordinary or partial differential equations, then this can be embedded into the objective function \cite{raissi2019physics}. %
Given a general system in the form of the PDE, with external input $\mathbf{p}$,
\begin{equation}
\mathcal{F}(\mathbf{y},\mathbf{x};\boldsymbol{\theta}_s) = \mathbf{p}
\end{equation}
for some nonlinear differential operator $\mathcal{F}$ acting on $\mathbf{y}(\mathbf{x})$, where $\boldsymbol{\theta}_s$ are parameters of the equation, and $\mathbf{p}$ is the acting external input. %

\begin{figure}[h!]
    \centering
    \includegraphics[width=\textwidth]{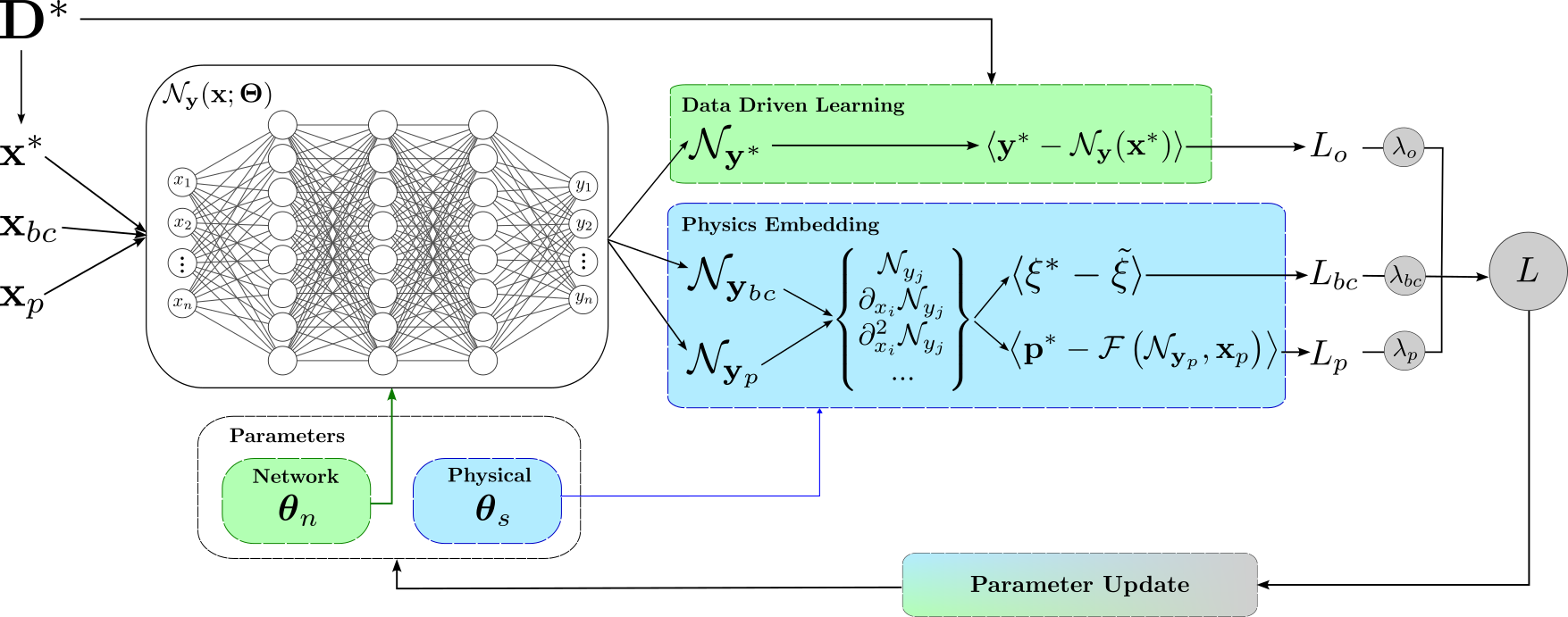}
    \caption{Framework of a general PINN, highlighting where the data-driven and physics-knowledge are embedded within the process}
    \label{fig:pinn_framework}
\end{figure}

\begin{figure}[h!]
    \centering
    \includegraphics[width=0.7\textwidth]{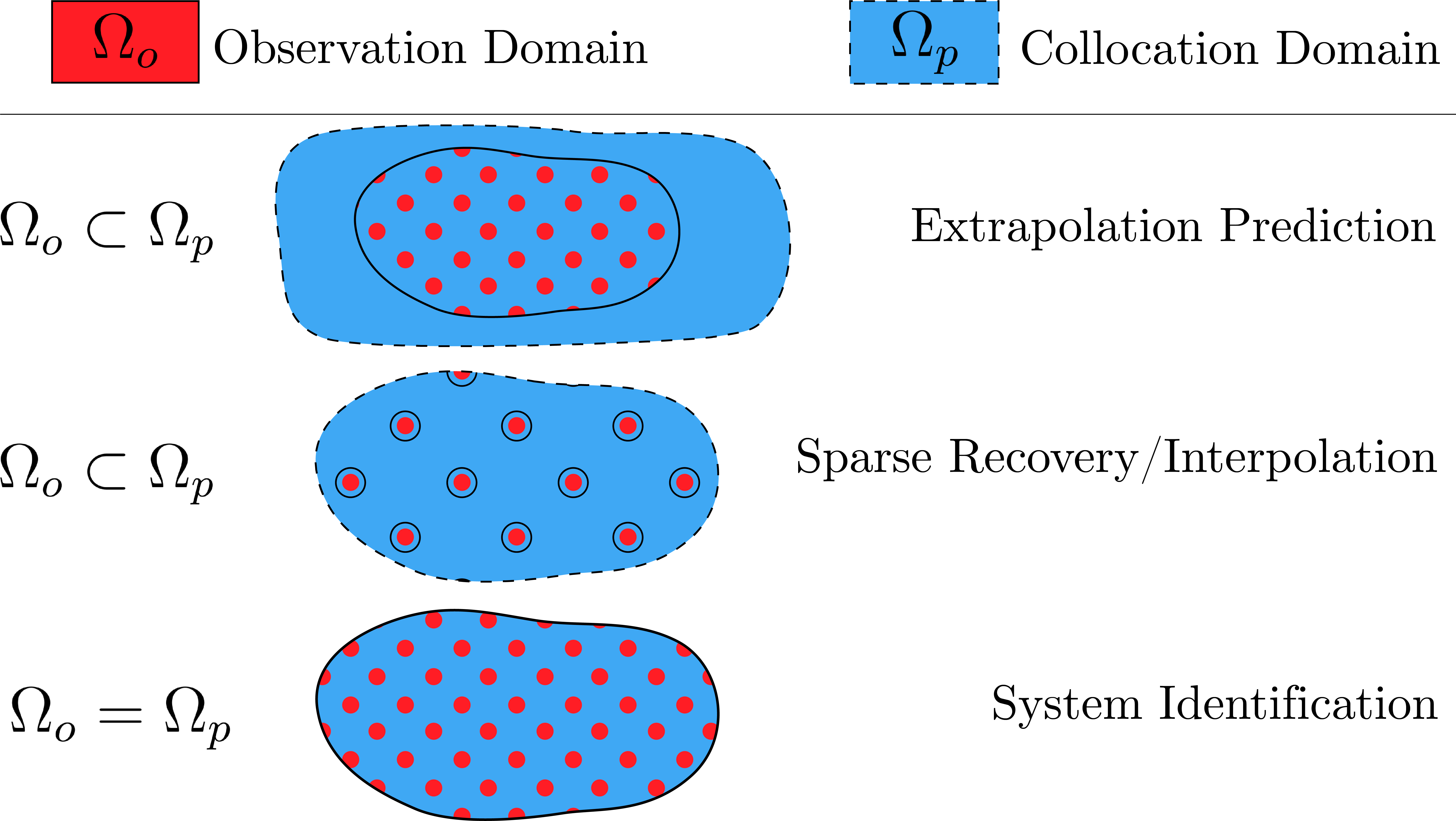}
    \caption{Visualisation of domain definitions for different schemes and motivations that employ PINNs. The blue areas represent the continuous collocation domain, and the red dots represent the coverage and sparsity of the discrete observation domain.}
    \label{fig:pinn_domains}
\end{figure}

The network prediction can then be used to estimate the result of the nonlinear operator, $\mathcal{F}(\mathcal{N}_{\mathbf{y}}(\mathbf{x};\boldsymbol{\theta}_n),\mathbf{x};\boldsymbol{\theta}_s) = \tilde{\mathbf{p}}$. %
The collocation and observation domains are important to distinguish; the observation domain is where measured target values are available and is a discrete domain, whereas the collocation domain is where the network aims to predict, and is continuous. %
A visualisation of this concept is provided in \Cref{fig:pinn_domains}. %
Given the domain of collocation points, $\Omega_p$, and observations of the system input, $\mathbf{p}^* \in \Omega_p$, an additional objective term is defined as,
\begin{equation}
    L_{p}(\mathbf{x};\boldsymbol{\theta}_n,\boldsymbol{\theta}_s) = \langle \mathbf{p}^* - \mathcal{F}(\mathcal{N}_\mathbf{y}(\mathbf{x};\boldsymbol{\theta}_n),\mathbf{x};\boldsymbol{\theta}_s) \rangle _{\Omega_p}
    \label{eq:physics_loss_gen}
\end{equation}
where the minimisation is based on the physics equation being satisfied when the value of $\langle\mathbf{p}^* - \tilde{\mathbf{p}}\rangle$ is zero, embedding an assumption that the differential operator $\mathcal{F}$ exhaustively captures the physics, i.e.\ there are no missing terms within the equation. %
Additionally, known boundary conditions can be embedded in a `soft' manner over the boundary domain $\partial\Omega \in \Omega_p$,
\begin{equation}
    L_{bc}(\mathbf{x};\boldsymbol{\theta}_n,\boldsymbol{\theta}_s) = \langle \xi^* - \xi(\mathcal{N}_{\mathbf{y}}(\mathbf{x};\boldsymbol{\theta}_n), \boldsymbol{\theta}_s) \rangle _{\partial\Omega}
\end{equation}

Then, a weighted sum of the individual loss terms forms the total loss,
\begin{equation}
    L = \lambda_o L_{o} + \lambda_p L_{p} + \lambda_{bc} L_{bc}
\end{equation}
where $\lambda_o, \lambda_p, \lambda_{bc}$ are normalisation parameters required to posit the loss terms in the same magnitude to aid optimisation, or weight their importance. %
On some occasions, some loss terms are omitted; e.g.\ when there are no observations, and the PINN is used as a forward modeller, only the boundary and physics loss terms are used. %

\subsection{Prescription for State-Space Systems}
\label{sec:pinn_dynam_prescr}

In this setup, the output of the network is a prediction of a given instance of the state $\mathbf{z}$ of a system undergoing a specific input (forcing) from prescribed initial conditions, using \emph{only} time as the network input,
\begin{equation}
    \tilde{\mathbf{z}}(t) = \mathcal{N}_{\mathbf{z}}(t;\boldsymbol{\theta}_w)
    \label{eq:state_est}
\end{equation}
In many practical setups, especially not in a laboratory setting, it is difficult to have direct observations of the state $\mathbf{z}$, and so direct implementation of \Cref{eq:obs_loss} would not be possible \emph{per se}. %
As they are the more practical and customary measurements, we here adopt the assumption of observations of acceleration, $\mathbf{y}^*(t) = \ddot{\mathbf{u}}^*(t)$, and force, $\mathbf{f}^*$, i.e.\ $\mathbf{S}_d = \mathbf{S}_v \in \boldsymbol{0}$. %
The estimated outputs are then formed from \Cref{eq:state_eq_output},
\begin{equation}
    \tilde{\mathbf{y}}(\mathbf{t};\boldsymbol{\theta}_w, \boldsymbol{\theta}_s)  = \mathbf{B}(\boldsymbol{\theta}_s)\mathcal{N}_{\mathbf{z}}(\mathbf{t};\boldsymbol{\theta}_w) + \mathbf{B}_n(\boldsymbol{\theta}_s) g_n\left(\mathcal{N}_{\mathbf{z}}(\mathbf{t};\boldsymbol{\theta}_w)\right) + \mathbf{D}(\boldsymbol{\theta}_s)\mathbf{f}^*(\mathbf{t})
    \label{eq:estimated_output}
\end{equation}
For implementation of the physics-loss, the differential operator is defined from \Cref{eq:governing_eq_state},
\begin{equation}
    \mathcal{F}(\mathcal{N}_{\mathbf{z}}(\mathbf{t};\boldsymbol{\theta}_w),\boldsymbol{\theta}_s) = \hat{\mathbf{H}}(\boldsymbol{\theta}_s)\left(\partial_t\mathcal{N}_{\mathbf{z}}(\mathbf{t};\boldsymbol{\theta}_w) - \mathbf{A}(\boldsymbol{\theta}_s)\mathcal{N}_{\mathbf{z}}(\mathbf{t};\boldsymbol{\theta}_w) - \mathbf{A}_n(\boldsymbol{\theta}_s) g_n\left(\mathcal{N}_{\mathbf{z}}(\mathbf{t};\boldsymbol{\theta}_w)\right) \right)
    \label{eq:pinn_diff_operator}
\end{equation}

\noindent where $\partial_t\mathcal{N}_{\mathbf{z}}(\mathbf{t};\boldsymbol{\theta}_w)$ is the estimated first order derivative of the state using automatic differentiation, and the quasi-inverted input matrix $\hat{\mathbf{H}}(\boldsymbol{\theta}_s)$ and estimated inputs $\tilde{\mathbf{p}}$ are, 
\begin{equation}
    \hat{\mathbf{H}}(\boldsymbol{\theta}_s) = \begin{bmatrix}
        \mathbf{I} & \boldsymbol{0} \\ \boldsymbol{0} & \mathbf{M}(\boldsymbol{\theta}_s)
    \end{bmatrix}, \quad
    \tilde{\mathbf{p}}(\mathbf{t}) = \tilde{\mathbf{f}}(\mathbf{t}) = \mathcal{F}(\mathcal{N}_{\mathbf{z}}(\mathbf{t};\boldsymbol{\theta}_w),\boldsymbol{\theta}_s) \left[ n:2n \right]
\end{equation}

The expansion of \Cref{eq:governing_eq_state,eq:pinn_diff_operator} provides a set of $2n$ equations, the first $n$ of which equate the velocity and the first derivative of displacement with respect to time. %
The latter $n$ equations equate the external and restoring forces of the system. %
From this, the boundary condition is instead posed as a `continuity condition', which imposes the equality established by the first $n$ equations, and the physics loss is established by the latter $n$ equations. %
The observation, continuity, and physics losses are then defined as,

\begin{subequations}
    \begin{equation}
        L_{o}(\mathbf{t};\boldsymbol{\theta}_w, \boldsymbol{\theta}_s) = \langle \mathbf{y}^* - \tilde{\mathbf{y}}(\mathbf{t};\boldsymbol{\theta}_w, \boldsymbol{\theta}_s) \rangle _{\Omega_o}
        \label{eq:obs_loss_ode}
    \end{equation}
    \begin{equation}
        L_{cc}(\mathbf{t};\boldsymbol{\theta}_w, \boldsymbol{\theta}_s) = \langle \mathcal{F}(\mathcal{N}_{\mathbf{z}}(\mathbf{t};\boldsymbol{\theta}_w),\boldsymbol{\theta}_s) \left[ 0:n \right] \rangle _{\Omega_p}
        \label{eq:contin_loss}
    \end{equation}
    \begin{equation}
        L_{p}(\mathbf{t};\boldsymbol{\theta}_w, \boldsymbol{\theta}_s) = \langle \mathbf{f}^* - \mathcal{F}(\mathcal{N}_{\mathbf{z}}(\mathbf{t};\boldsymbol{\theta}_w),\boldsymbol{\theta}_s) \left[ n:2n \right] \rangle _{\Omega_p}
        \label{eq:physics_loss}
    \end{equation}
    \label{eq:all_losses}
\end{subequations}

An overview of the general algorithm for applying a PINN to a MDOF oscillator state-space model is shown in \Cref{alg:state-space-pinn}. %
The algorithm is designed to predict the state of the system over the specified collocation domain, and optionally the system parameters, by minimising the loss functions defined in \Cref{eq:all_losses}. %
This is done by iteratively updating the network parameters $\boldsymbol{\theta}_n$ and system parameters $\boldsymbol{\theta}_s$ using the gradients of the loss functions. %
Construction, performance, and deployment of such an algorithm requires the use of a deep learning framework with autodifferentation capabilities, such as TensorFlow \cite{tensorflow2015-whitepaper} or PyTorch \cite{paszke_automatic_2017}, the latter of which is used in this work for the implementation of the PINN algorithm. %

\begin{algorithm}
\caption{PINN algorithm for state estimation and optional parameter estimation for observer state space model.}
\label{alg:state-space-pinn}
\begin{algorithmic}[1] 
    \State \textbf{Input:} Observation Data $\mathbf{D}^*$, Collocation Data
    \State \textbf{Output:} Predictions of State Over Collocation Domain, Predictions of System Parameters
    \State $\mathbf{t}^*, \ddot{\mathbf{u}}^*, \mathbf{f}^* \gets \mathbf{D}^*$
    \State $\mathbf{t}_c \gets \mathbf{D}$
    \State Initialise neural network $\mathcal{N}(t;\boldsymbol{\theta}_n)$
    \State \textit{Optional:} Initialise system parameters to be estimated $\boldsymbol{\Theta}_s$
    \For{$i = 0$ to epochs}
        \State $\tilde{\mathbf{z}}_{\Omega_o} \gets \mathcal{N}(\mathbf{t}^*;\boldsymbol{\theta_n})$ \Comment{Predict states in observation domain}
        \State $\tilde{\mathbf{z}}_{\Omega_p} \gets \mathcal{N}(\mathbf{t}_c;\boldsymbol{\theta_n})$ \Comment{Predict states in collocation domain}
        \State $\partial_t \tilde{\mathbf{z}}_{\Omega_p} \gets \texttt{grad}(\tilde{\mathbf{z}}, \mathbf{t}_c)$ \Comment{Retrieve gradient in collocation domain}
        \If{$\boldsymbol{\Theta_s} \neq \emptyset$}
            \State $\boldsymbol{\theta}_s \gets \boldsymbol{\Theta}_s$ \Comment{Populate system parameters}
        \EndIf
        \State $\tilde{\mathbf{f}}_{\Omega_p} \gets \mathcal{F}(\tilde{\mathbf{z}}_{\Omega_p}, \partial_t \tilde{\mathbf{z}}_{\Omega_p}, \boldsymbol{\Theta_s})$ \Comment{Predict forces in collocation domain}
        \State $L \gets L_o(\tilde{\mathbf{z}}, \mathbf{z}^*), L_{cc}, L_p(\tilde{\mathbf{f}}, \mathbf{f}^*)$ \Comment{See loss functions in reference equations}
        \State Update $\boldsymbol{\theta_n}$ using $\nabla L$
        \If{$\boldsymbol{\Theta_s} \neq \emptyset$}
            \State Update $\boldsymbol{\Theta_s}$ using $\nabla L$
        \EndIf
    \EndFor
\end{algorithmic}
\end{algorithm}

\section{Case Study Descriptions}

When applying the PINN framework outlined in \Cref{sec:pinn_dynam_prescr}, a number of different scenarios and motivations are here constructed as case studies with the intention of demonstrating use of PINNs for the identified downstream tasks, namely sparse recovery, state-parameter estimation, and parameter estimation of dynamical systems. %
As previously mentioned, the use of the PINN requires the use of a physics equation which is assumed to exhaustively capture the physics, and so, the first two of these schemes have been further split into three cases corresponding to modes of inaccuracy in the prescribed model;
\begin{enumerate}
    \item \emph{No model uncertainty}; the assumed known model form is the true system model.
    \item \emph{Model form uncertainty}; the assumed known model form is assumed to be the equivalent linear system (i.e.\ zero nonlinear forces in \Cref{eq:governing_eq_nom}).
    \item \emph{Model parameter uncertainty}; the parameters of the true system are randomly varied by $\pm 5\textrm{-}10\%$ and the assumed known parameters are kept at the nominal values.
\end{enumerate}

\subsection{State Estimation (Sparse Recovery)}
For this scheme, the aim is to predict the state \emph{outside} the domain of observations, $\Omega_o \subset \Omega_p$, and is performed in a domain extension context. %
Here, sparsity of the observation domain is meant in the spatial sense, i.e., in terms of the degrees of freedom that are assumed to be monitored, as this is a more useful and applicable context for application as opposed to the alternate assumption of reducing observations in the time domain (e.g. missing samples). This reduction in the observation domain is specified in terms of the percentage of unobserved degrees of freedom $b_r$. %
Thus, the observation domain forms a subset of the complete set of $n$ degrees of freedom, practically corresponding to sparse sensor locations, where $d<n$,
\begin{equation}
    \mathbf{y}^* = \{\mathbf{y}^{(i)}\} \; \textrm{for } i \textrm{ in } \mathbf{d}=\{i_1, i_2, ... , i_d\}
\end{equation}
This dictates the formation of the selection matrix of the acceleration observations,
\begin{equation}
    \mathbf{S}_a = \textrm{diag}(\mathbf{d})
\end{equation}
For domain extension, the scope of the domain of observations is reduced by removing the last $n_{r}$ degrees of freedom. %
The number of unobserved degrees of freedom is calculated by $n_r = \textrm{round}(b_r n)$ and is related to the subset indexes by $\mathbf{d} = \{1, 2, ... , n-n_r+1, n-n_r\}$. %


\subsection{State-Parameter Estimation}

In addition to estimating unobserved states of the signal, in this scheme, unknown systems parameters are also estimated. %
In order to maintain observability and identifiability of the system, for the state-parameter estimation regime the observed DOFs are reduced by 40\% by having $\mathbf{d} =\{2, 4\}$. %
Then, the linear and nonlinear stiffness connections represented by $\mathbf{d}$ are also assumed unknown and form the set of learnable parameters $\boldsymbol{\Theta}_s = \boldsymbol{\theta}_s[\mathbf{d}]$. %

\subsection{Parameter Estimation}
When exploiting the PINN architecture solely for parameter estimation, the collocation and observation domain are equal, $\Omega_p = \Omega_o$. %
The setup of the loss function \Cref{eq:all_losses} will provide a deterministic estimate of the state and the physical system parameters. %
Here, we propose an alternative objective function prescription based on a Gaussian likelihood. %
We consider the observed values of the state and force as processes with mean $\bar{\mathbf{z}}$ and random Gaussian noise;
\begin{subequations}
    \begin{equation}
        \mathbf{z}^* = \bar{\mathbf{z}} + \varepsilon_z, \quad \varepsilon_z \sim \textrm{Normal}(0, \boldsymbol{\Sigma}_{\mathbf{z}})
    \end{equation}
    \begin{equation}
        \mathbf{f}^* = \bar{\mathbf{f}} + \varepsilon_f, \quad \varepsilon_f \sim \textrm{Normal}(0, \sigma_f)
    \end{equation}
\end{subequations}
where $\boldsymbol{\Sigma}_{\mathbf{z}}$ is populated by assuming the noise distribution to be consistent for each state variable and independent of the degree of freedom. %
Thus, it is diagonal and populated by $\sigma_u$ and $\sigma_{\dot{u}}$, representing the variance for the displacement and velocity, respectively. %
\begin{equation}
    \boldsymbol{\Sigma}_{\mathbf{z}} \in \mathbb{R}^{2n\times 2n} = \begin{bmatrix}
        \sigma_u\mathbb{I}^{n} & \boldsymbol{0} \\ \boldsymbol{0} & \sigma_{\dot{u}}\mathbb{I}^{n}
    \end{bmatrix}
\end{equation}
Then, using \Cref{eq:state_est,eq:pinn_diff_operator}, the observation and physics likelihoods become,
\begin{subequations}
    \begin{equation}
        p\left( \mathbf{z}^*|\boldsymbol{\theta}_n \right) = \left( 2\pi \right)^{-n}|\boldsymbol{\Sigma}_{\mathbf{z}}|^{-1/2} \prod^{N_o}_{i=1} \exp\left[ -\frac{1}{2}\left(\mathbf{z}^*_i-\mathcal{N}_{\mathbf{z}}(t^{(i)};\boldsymbol{\theta}_n)\right)^T \boldsymbol{\Sigma}_{\mathbf{z}}^{-1} \left(\mathbf{z}^*_i-\mathcal{N}_{\mathbf{z}}(t^{(i)};\boldsymbol{\theta}_n)\right) \right]
        \label{eq:obs_likeli}
    \end{equation}
    \begin{equation}
        p\left( \mathbf{f}^*|\boldsymbol{\theta}_s, \boldsymbol{\theta}_n \right) = \prod^{N_c}_{i=1} \frac{1}{\sqrt{2\pi\sigma_f^2}} \exp \left( -\frac{1}{2} \frac{||\mathbf{f}^*_i-\mathcal{F}(\mathcal{N}_{\mathbf{z}}(t^{(i)};\boldsymbol{\theta}_n),\boldsymbol{\theta}_s) \left[ n:2n \right]||^2}{\sigma_f^2} \right)
        \label{eq:physics_likeli}
    \end{equation}
\end{subequations}
For optimisation, the objective functions are formulated as the negative log-likelihood,
\begin{equation}
    L_o = -\log \left( p\left(\mathbf{z}^*|\boldsymbol{\theta}_n\right) \right), \qquad
    L_p = -\log \left( p\left(\mathbf{f}^*|\boldsymbol{\theta}_s,\boldsymbol{\theta}_n\right) \right)
\end{equation}
and the noise parameters are included within the parameter set to be optimised over. %

At this stage, the motivation is to estimate the posterior distribution of the system parameters $\boldsymbol{\theta}_s$, given a set of observed values of the input force process and estimated mean values of the state, $p(\boldsymbol{\theta}_s|\mathbf{f}^*, \boldsymbol{\theta}_n)$. %
This is not directly inferrable, but can be posited using a manipulation of \Cref{eq:obs_likeli,eq:physics_likeli} and Bayes' rule,
\begin{equation}
    p(\theta_s|\mathbf{f}^*, \boldsymbol{\theta}_n) = \frac{p(\mathbf{f}^*|\boldsymbol{\theta}_s, \boldsymbol{\theta}_n)p(\boldsymbol{\theta}_s|\boldsymbol{\theta}_n)}{\int_{\Omega\boldsymbol{\theta}_s}p(\mathbf{f}^*|\boldsymbol{\theta}_n, \boldsymbol{\theta}_s)p(\boldsymbol{\theta}_s| \boldsymbol{\theta}_n)d\boldsymbol{\theta}_s}
\end{equation}
where $p(\mathbf{f}^*|\boldsymbol{\theta}_s, \boldsymbol{\theta}_n)$ is calculated using \Cref{eq:physics_likeli}, and the prior over the system parameters is assumed independent of the network parameters ($p(\boldsymbol{\theta}_s|\boldsymbol{\theta}_n)=p(\boldsymbol{\theta}_s)$). 
A solution for this could be proposed by prescribing a posterior distribution form (e.g.\ Gaussian), and returning deterministic estimates of the parameters governing these distributions (e.g.\ $\sigma_k$). %
However, the problem is then not solved, and is instead now the denominator $\int_{\Omega\boldsymbol{\theta}_s}p(\mathbf{f}^*|\boldsymbol{\theta}_s, \boldsymbol{\theta}_n)p(\boldsymbol{\theta}_s)d\boldsymbol{\theta}_s$ is intractable. %
Instead here, MCMC sampling is used, which samples from the posterior with enough repetition that an estimate of the posterior distribution over the parameters can be inferred. %
This allows for freedom in the posterior shape, as well as capturing any multivariate correlation between the parameters. %
An outline of the derivation and procedure for MCMC is given in \cite{barber_bayesian_2012}, and here the NUTS algorithm \cite{hoffman_no-u-turn_2014}, which exploits Hamiltonian dynamics, is employed. %
For this work, the PyTorch-based Pyro \cite{bingham_pyro_2019} package is used to perform the sampling procedure. %



\section{Case Studies}\label{sec:results}

The results of the case studies are presented below, grouped into the various schemes and in each scheme on the different case scenarios. %
When prediction results are shown, the first two states ($\mathbf{u}$ and $\dot{\mathbf{u}}$) are the direct outputs from the neural network $\mathcal{N}_{\mathbf{z}}$, and the acceleration and force predictions are calculated using \Cref{eq:estimated_output} and \Cref{eq:pinn_diff_operator}, respectively. %

\subsection{Sparse Recovery}

In the first scheme, when observations are available, only measurements of the acceleration $\ddot{\mathbf{u}}^*$ and force $\mathbf{f}^*$ are provided. %
For all cases, the unobserved degrees of freedom represent the latter 67\% of the DOFs of the system, and the observation domain is the first 33\% of the DOFs. %


\begin{figure}[H]
    \centering
    \includegraphics[width=\linewidth]{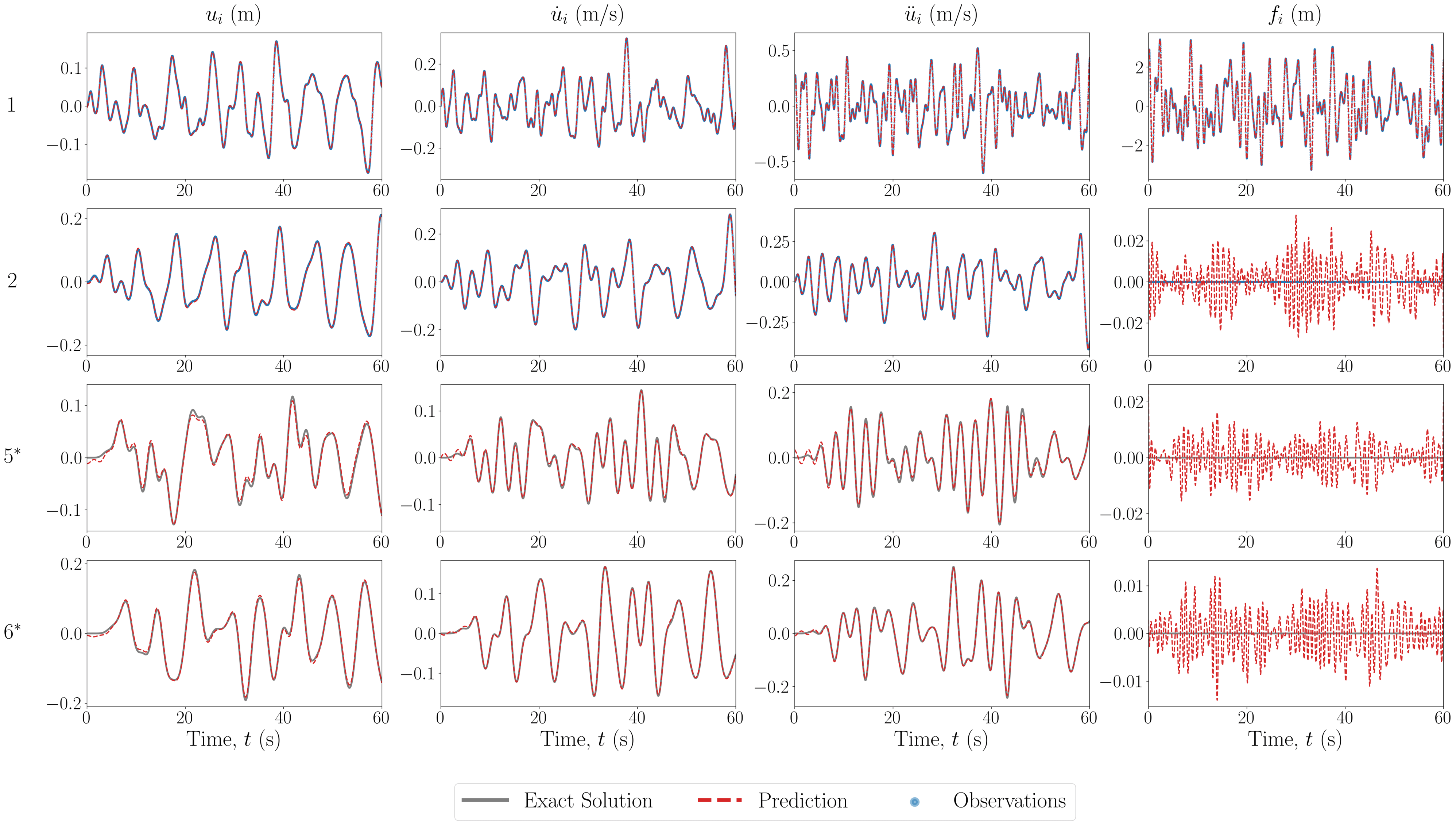}
    \caption{Estimated input and response for 6DOF Duffing system with no model error. The DOF is indicated on the left, with an asterisk indicating unobserved DOFs.}
    \label{fig:sr-nonerr-duf-6dof}
\end{figure}

\begin{figure}[H]
    \centering
    \includegraphics[width=\linewidth]{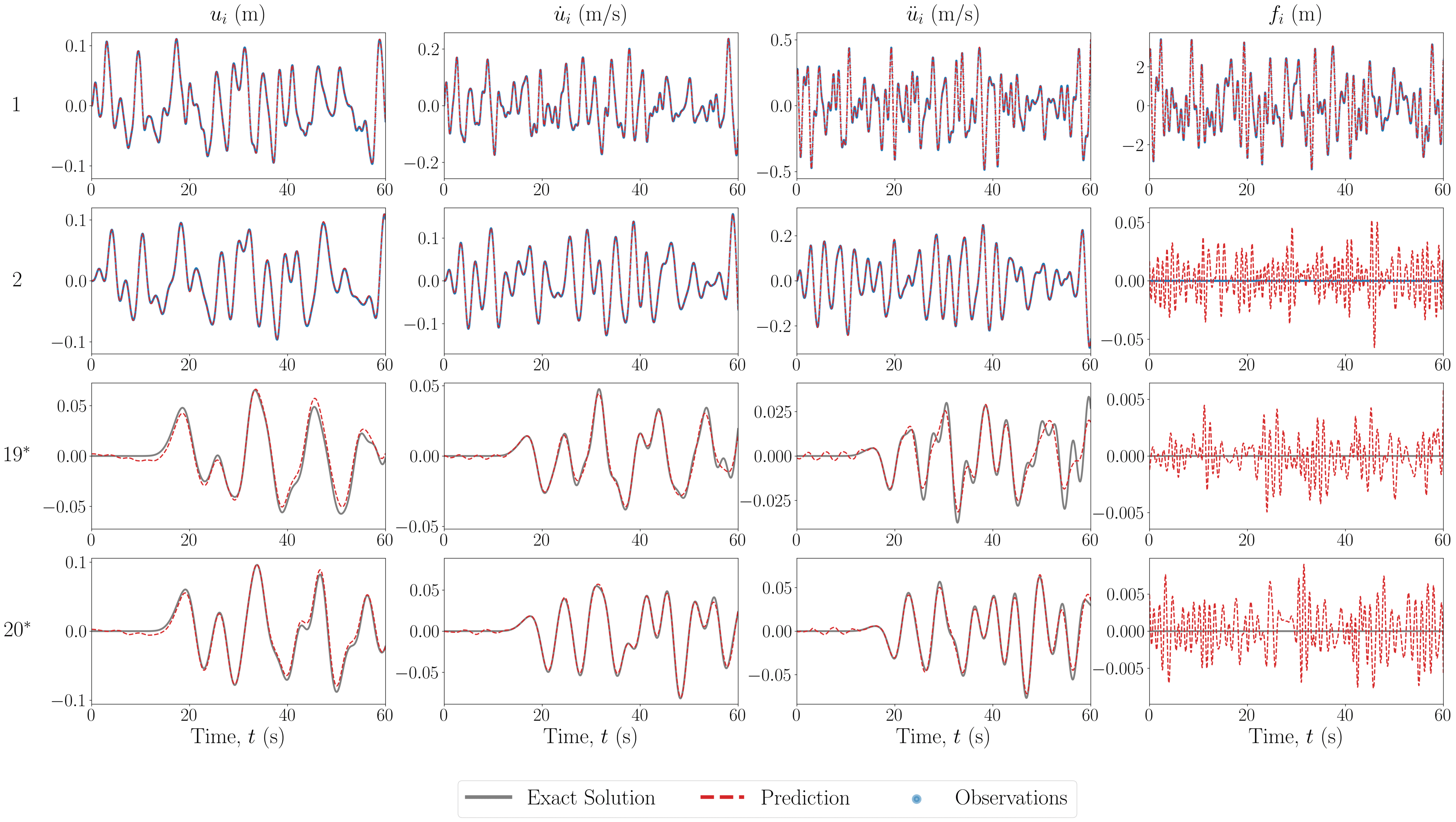}
    \caption{Estimated input and response for 20DOF Duffing system with no model error. The DOF is indicated on the left, with an asterisk indicating unobserved DOFs.}
    \label{fig:sr-nonerr-duf-20dof}
\end{figure}

The first case performed is that of sparse recovery with no errors in the model which is prescribed for the PINN. %
The results of the state estimation for the first and last two degrees of freedom are shown in \Cref{fig:sr-nonerr-duf-6dof} for the 6DOF Duffing system, with the unobserved degrees of freedom noted by an asterisk. %
The estimated recovered states match well with that of the ground truth, even at the DOFs furthest away from those which are available in the measurements. %
The estimated forces also match well at the observed DOFs, and are small when estimating at DOFs where zero forces are present, in comparison to the magnitude of the input force. %
They appear to be following a pattern or signal that is not necessarily noise, indicating that most of the estimation error comes as an artefact by using the predicted states to predict the force. %

\begin{figure}[h!]
    \centering
    \includegraphics[width=\linewidth]{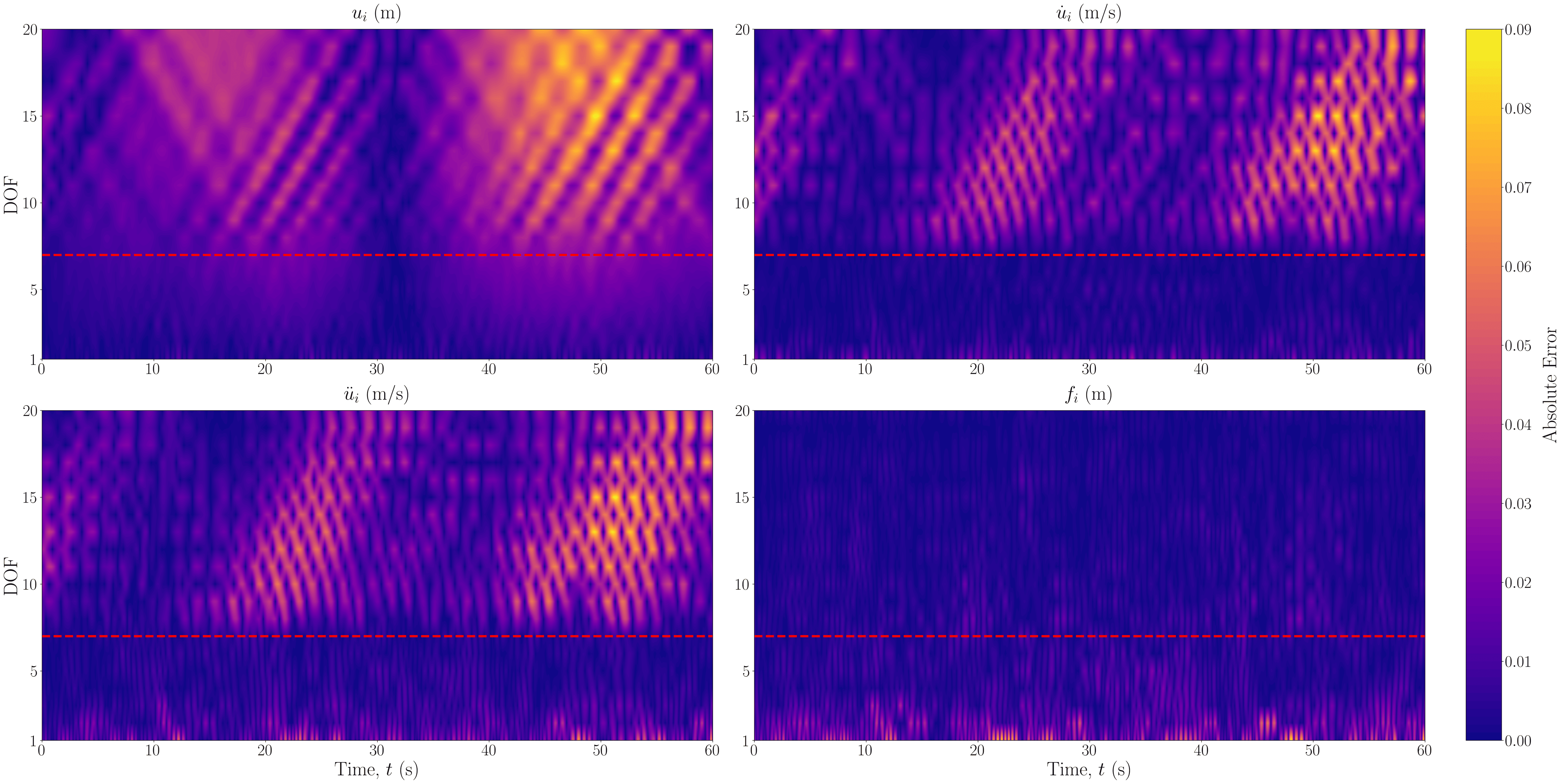}
    \caption{Absolute errors between predicted states and forces and the ground truth for Duffing 20DOF system with no error of the model prescribed in the PINN. The red dashed line represents the boundary of the observation domain.}
    \label{fig:sr-20dof-errmat}
\end{figure}

Next, the scheme was also tested with a 20-DOF Duffing oscillator, the results of which are shown in \Cref{fig:sr-nonerr-duf-20dof}. %
From this figure, it appears that the predictions further away from the area of observations begin to worsen. %
To study the propagation and level of error throughout the system, the absolute error between the predictions and the ground truth over all DOFs is also shown in \Cref{fig:sr-20dof-errmat}, where the red dashed line represents the boundary of the observation domain. %
It can be seen that the prediction does not necessarily worsen with distance from the observation domain, but rather the error is more pronounced in paths of vibration mode propagation. %
This could be as a result of the superposition of a larger number of vibration modes, and thus the objective function minimum is in a more complex optimisation space. %
Furthermore, the magnitude of the error in the forcing signal is small, and mostly exists in the recovery of the forcing signal at the observed DOFs.

\begin{figure}[h!]
    \centering
    \includegraphics[width=\linewidth]{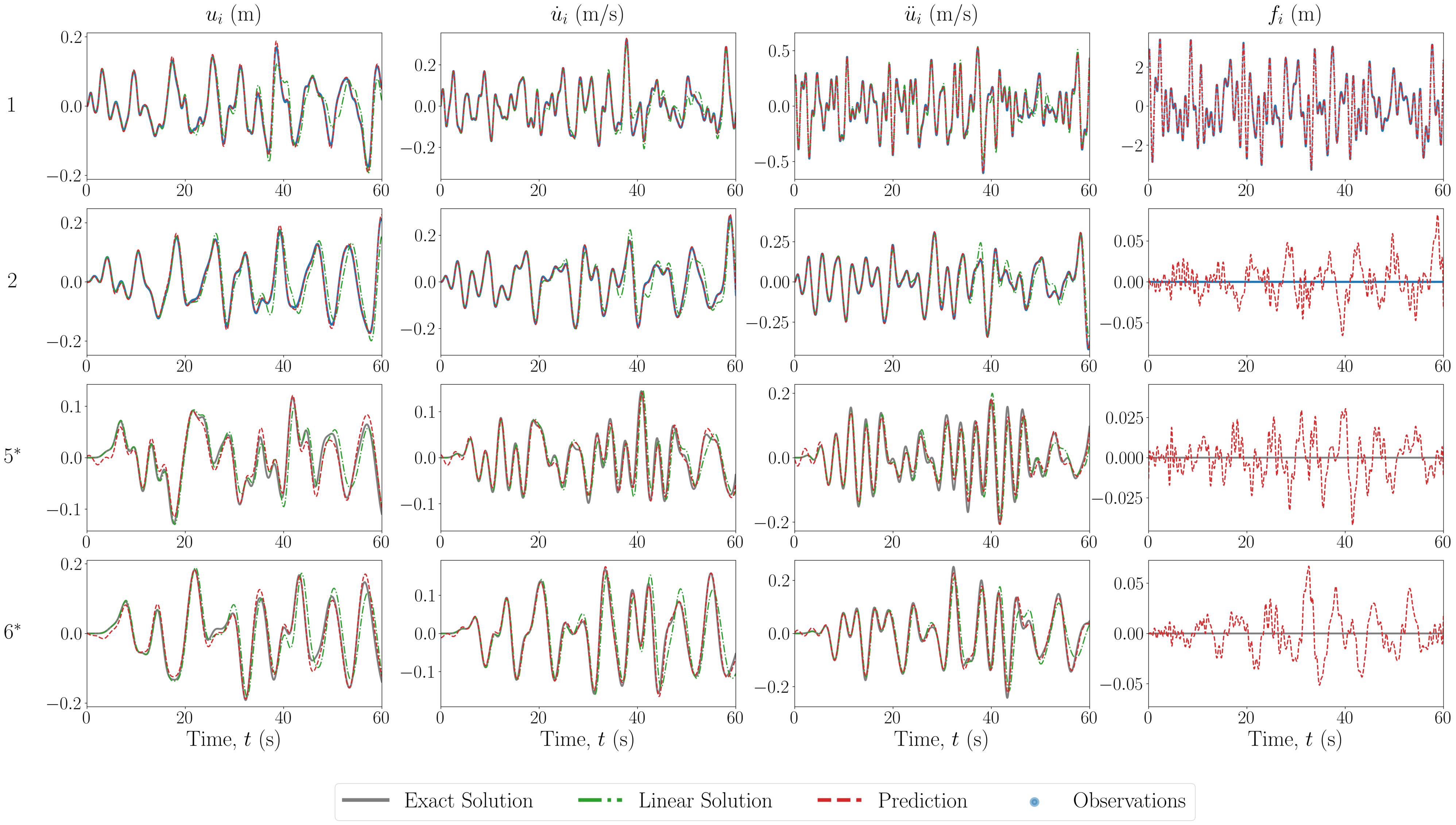}
    \caption{Estimated input and response for 6DOF Duffing system with model error in the form of a prescribed linear model. The blue dash-dot line represents the true solution of the equivalent linear system under the same forcing. The DOF is indicated on the left, with an asterisk indicating unobserved DOFs.}
    \label{fig:sr-linmod-duf-6dof}
\end{figure}

In the next case scenario, the model prescribed in the PINN is set to be the equivalent linear system, i.e.\ the nonlinear forces are set to zero. %
The results of the state estimation for the 6DOF Duffing system are shown in \Cref{fig:sr-linmod-duf-6dof}. %
As well as the true nonlinear system results, the simulated response of the equivalent linear system under the same forcing is shown in the blue dash-dot line. %
It can be seen that the predictions of the states are not as accurate as in the case with no model errors, and the predictions of the states further away from the observation domain are less accurate. %
However, this error is still relatively minimal, especially when compared to the estimated response of the equivalent linear system. %
This could be a result of the objective function minimisation only being relative to each instance time, and so the error from the assumed zero nonlinear restoring forces does not propagate throughout the time span. %
Once again, the recovered force is matches closely with that of the ground truth, however, for the case of the linear system prescription, the recovered force at locations where it should be zero appears to be more structured than in previous cases. %
This could be explained as artefacts from the model not having captured all of the restoring forces in the system, i.e. the nonlinear restoring forces. %


\begin{figure}[h!]
    \centering
    \includegraphics[width=\linewidth]{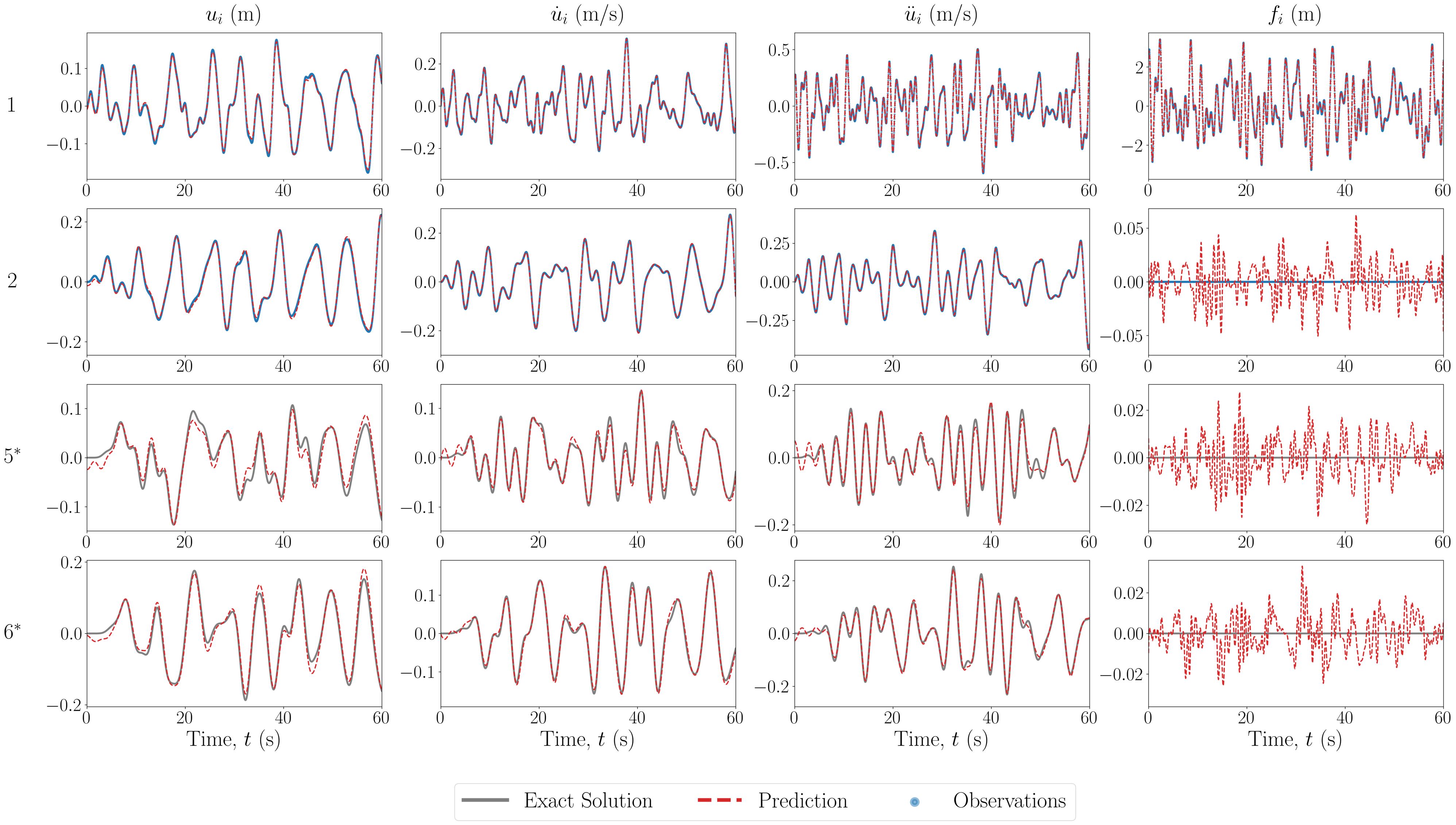}
    \caption{Estimated input and response for 6DOF Duffing system with model error in the form of inaccurate system parameter values. The DOF is indicated on the left, with an asterisk indicating unobserved DOFs.}
    \label{fig:sr-valerr-duf-6dof}
\end{figure}

In the final sparse recovery case scenario, the model error is of the form of system parameter value discrepancies. %
The results of the state estimation for this case applied to a 6DOF Duffing oscillator are shown in \Cref{fig:sr-valerr-duf-6dof}. %
The values of the true system used for simulation, along with the prescribed values, are shown in \Cref{tab:sr-valerr-duf-prescr}. %
Similarly to the case with no model errors, the predictions of the states are accurate at the observed degrees of freedom. %
However, there remains some errors, naturally, in the predictions further away from the observation domain. %
This differs from when the model error is in the form of a prescribed linear system, as the errors only become more pronounced \emph{outside} the domain of observation. %
As with the schemes with no modelling errors, the recovery of the forcing signals are very small in comparison to magnitude of the input forces, and there apparent signal pattern is likely an artefact from the state estimation. %

\begin{table}[h!]
    \caption{Values of prescribed and true parameters applied to Duffing 6DOF system with model prescription errors in the form of inaccurate system parameter values.}
    \label{tab:sr-valerr-duf-prescr}
    \begin{tabular}{rcccccc}
        & \multicolumn{6}{@{}c@{}}{$\mathbf{k}$} \\
        \cmidrule{2-7}
       & $k_1$ & $k_2$ & $k_3$ & $k_4$ & $k_5$ & $k_6$ \\
       \midrule
       Prescribed & 15.0 & 15.0 & 15.0 & 15.0 & 15.0 & 15.0\\
       True       & 13.83&	13.91&	13.89&	16.05&	15.76&	16.39\\
       \midrule
        & \multicolumn{6}{@{}c@{}}{$\mathbf{c}$} \\
        \cmidrule{2-7}
       & $c_1$ & $c_2$ & $c_3$ & $c_4$ & $c_5$ & $c_6$ \\
       \midrule
       Prescribed       & 1.0 & 1.0 & 1.0 & 1.0 & 1.0 & 1.0\\
       True  & 0.92&	0.93&	0.93&	1.07&	1.05&	1.09\\
       \midrule
        & \multicolumn{6}{@{}c@{}}{$\boldsymbol{\kappa}$} \\
        \cmidrule{2-7}
       & $\kappa_1$ & $\kappa_2$ & $\kappa_3$ & $\kappa_4$ & $\kappa_5$ & $\kappa_6$ \\
       \midrule
       Prescribed & 100.0 & 100.0 & 100.0 & 100.0 & 100.0 & 100.0 \\
       True       & 92.23 & 92.75 &	92.63&	106.99&	105.06&	109.28 \\
       \botrule
    \end{tabular}
\end{table}

\Cref{tab:sr-summary} summarises the fitness metrics of the case studies performed for the sparse recovery scheme. %
The fitness is calculated by 
\begin{equation}
    \textrm{Fitness} = \left(1 - \frac{||\mathbf{x} - \tilde{\mathbf{x}}||}{||\mathbf{x} - \bar{\mathbf{x}}||} \right) \times 100
    \label{eq:fitness}
\end{equation}
where $\mathbf{x}$ is the ground truth, $\tilde{\mathbf{x}}$ is the prediction, and $\bar{\mathbf{x}}$ is the mean of the ground truth. %
From \Cref{tab:sr-summary}, it can be seen that the fitness of the predictions is generally high, with the lowest fitness being 95.97\% for the case with model error in the form of inaccurate system parameter values. %
When comparing the fitness of the predictions of the first and last degrees of freedom, it can be seen that the fitness of the predictions of the first degrees of freedom is generally higher than that of the last degrees of freedom. %
Furthermore, the fitness of the acceleration of the 20DOF system with no error is similarly worse than for the 6DOF systems with errors. %
This appears to indicate that the extent to which one is attempting to extend the domain of prediction has a similar impact on the accuracy of prediction as having an inaccurate model prescription. %
However, for the estimation of the displacement, the fitness is generally worse when there are model errors, particularly when the values of the system parameters are inaccurate. %
The modelling errors do not appear to have a significant impact on the accuracy of the force signal estimation, which is instead worsened when a larger level of noise is apparent, reducing the fitness to 94\%, whereas all other cases are above 99\% for fitness of the recovered force. %

\begin{table}[h!]
    \caption{Summary of fitness metrics in percent (\%) of the case studies performed for the sparse recovery scheme.}\label{tab:sr-summary}
    \centering
    \begin{tabular*}{\textwidth}{@{\extracolsep\fill}ccccccccccccc}
    \toprule
             & \multicolumn{4}{@{}c@{}}{All} & \multicolumn{4}{@{}c@{}}{First DOF} & \multicolumn{4}{@{}c@{}}{Last DOF} \\\cmidrule{2-5}\cmidrule{6-9}\cmidrule{10-13}
        Case & $\mathbf{u}$ & $\dot{\mathbf{u}}$ & $\ddot{\mathbf{u}}$ & $\mathbf{f}$ & $u_1$ & $\dot{u}_1$ & $\ddot{u}_1$ & $f_1$ & $u_n$ & $\dot{u}_n$ & $\ddot{u}_n$ & $f_n$ \\
        \midrule
        \begin{tabular}[c]{c}6DOF\\No Error\\SNR: 50.0\end{tabular} & 98.57 & 99.30 & 99.11 & - & 99.59 & 99.79 & 99.74 & 99.74 & 98.42 & 99.28 & 99.28 & -  \\
        \midrule
        \begin{tabular}[c]{c}6DOF\\No Error\\SNR: 20.0\end{tabular} & 99.13 & 99.46 & 99.32 & - & 99.56 & 99.66 & 99.58 & 99.49 & 99.10 & 99.50 & 99.46 & - \\
        \midrule
        \begin{tabular}[c]{c}6DOF\\No Error\\SNR: 10.0\end{tabular} & 95.47 & 97.89 & 97.59 & - & 97.60 & 98.35 & 96.48 & 94.07 & 95.25 & 97.79 & 98.40 & - \\
        \midrule
        \begin{tabular}[c]{c}20DOF\\No Error\\SNR: 50.0\end{tabular} & 97.74 & 98.70 & 97.73 & - & 99.65 & 99.74 & 99.61 & 99.62 & 97.14 & 98.47 & 97.46 & - \\
        \midrule
        \begin{tabular}[c]{c}6DOF\\Linear Model\\SNR: 50.0\end{tabular} & 96.60 & 97.95 & 97.64 & - & 98.51 & 99.51 & 99.57 & 99.49 & 96.19 & 97.60 & 97.44 & - \\
        \midrule
        \begin{tabular}[c]{c}6DOF\\Value Error\\SNR: 50.0\end{tabular} & 96.30 & 98.06 & 97.94 & - & 98.77 & 99.64 & 99.56 & 99.46 & 95.97 & 97.72 & 97.72 & - \\
    \botrule
    \end{tabular*}
    \footnotetext{Note: No fitness score is available for some force vector estimates as the true signal is zero in these locations (unforced).}
\end{table}

\subsection{State-Parameter Estimation}

In the second scheme, the objective is to generate predictions of both states at measured and unmeasured degrees of freedom, and estimate unknown physical system parameters. %
In the second scheme, for all cases, the unobserved degrees of freedom are the second and fourth DOF, which are in the form of measured accelerations $\ddot{\mathbf{u}}^*$ and forces $\mathbf{f}^*$. %
The parameters to determine are set as a subset of the total parameters $\boldsymbol{\Theta}_s \subset \boldsymbol{\theta}_s$, and are the stiffness and nonlinear components of the second and fourth connection, e.g.\ $\boldsymbol{\Theta}_s=\{k_2,k_4,\kappa_2,\kappa_4\}$ for a Duffing system. %

\begin{figure}[h!]
    \centering
    \includegraphics[width=\linewidth]{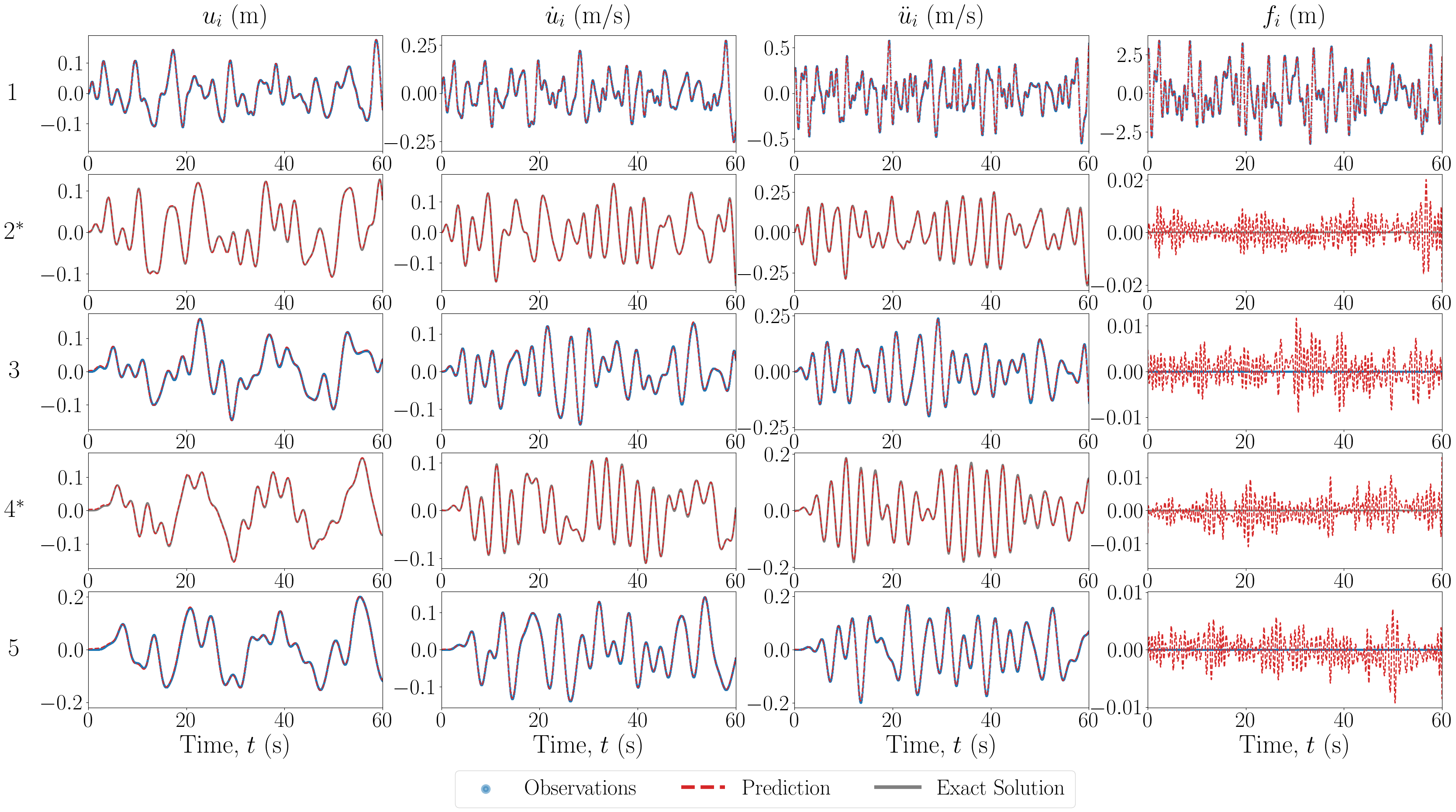}
    \caption{Estimated input and response for 5DOF Duffing system, from the state-parameter estimation scheme, with no model error. The DOF is indicated on the left, with an asterisk indicating unobserved DOFs.}
    \label{fig:spe-nonerr-duf}
\end{figure}

In the first case for this scheme, the model prescription used in the PINN is assumed to correspond to no model error, with the results of the state and force estimation shown in \Cref{fig:spe-nonerr-duf} for a 5DOF Duffing system. %
The estimated values of the system parameters are shown in \Cref{tab:spe-nonerr-duf}. %

The response and force prediction matches well with that of the ground truth, even for unobserved degrees of freedom. %
The estimated values of the system parameters are also close to the true values, with the largest error being 6\% for $\kappa_2$. %
This larger error could be a result of the relative restoring force being more sensitive to the displacement. %
This is especially significant in the fact that the optimisation is only relative to each instance time, which although gave an advantage in the estimation of response when there is a model error, may here provide a disadvantage in that no ``memory'' is used to continuously update the parameter estimation. %

\begin{table}[h!]
    \caption{Values of estimated parameters from state-parameter estimation scheme applied to Duffing MDOF system with no error in model form prescribed in the PINN.}\label{tab:spe-nonerr-duf}
    \begin{tabular*}{\textwidth}{@{\extracolsep\fill}lccccccc}
        \toprule
        & & \multicolumn{2}{@{}c@{}}{SNR: 50.0} & \multicolumn{2}{@{}c@{}}{SNR: 20.0} & \multicolumn{2}{@{}c@{}}{SNR: 10.0} \\
        \cmidrule{3-4}\cmidrule{5-6}\cmidrule{7-8}
        Parameter & True Value & Estimated & Error & Estimated & Error & Estimated & Error \\
        \midrule
       $\Theta_s^{(1)}=k_2$ & 15.0 & 14.91 & 0.615\% & 14.83 & 1.16\% & 15.02 & 0.145\% \\
       $\Theta_s^{(2)}=k_4$ & 15.0 & 14.95 & 0.370\% & 14.84 & 1.09\% & 14.86 & 0.931\% \\
       $\Theta_s^{(3)}=\kappa_2$ & 100.0 & 106.0 & 6.00\% & 84.32 & 15.68\% & 91.01 & 8.99\% \\
       $\Theta_s^{(4)}=\kappa_4$ & 100.0 & 99.674 & 0.326\% & 102.36 & 2.36\% & 69.16 & 30.84\% \\
       \botrule
    \end{tabular*}
\end{table}

To assess the effect of noise on the scheme, the observation signals were corrupted with noise representing a SNR of 20.0 and 10.0. 
The estimated values of the system parameters are shown also in \Cref{tab:spe-nonerr-duf}. %
The estimated reproduced state signals match well with those of the exact non-noisy signal, indicating the capability of the physics-information to generate an appropriate estimate of the mean, given a single repetition of the signal. %

The estimation of the system parameters appears to have been negatively affected by the larger level of noise corruption, which is not a surprising result. %
In particular, the nonlinear stiffness parameters return significantly worse predictions. %
This could be appropriated to the fact that the Duffing stiffness is much more sensitive to inaccuracies in the displacement estimation, and as seen in \Cref{tab:spe-summary}, the noisier signals return a lower fitness score. %
Furthermore, the error for $\kappa_2$ is larger than $\kappa_4$ when the SNR is 20.0 and the inverse is true when the SNR is 10.0. which may indicate the method's sensitivity to the initial conditions and optimisation process. %
However, overall, the estimated parameters are still reasonably well obtained, even with a single observation of the noisy signal. %

\begin{figure}[h!]
    \centering
    \includegraphics[width=\linewidth]{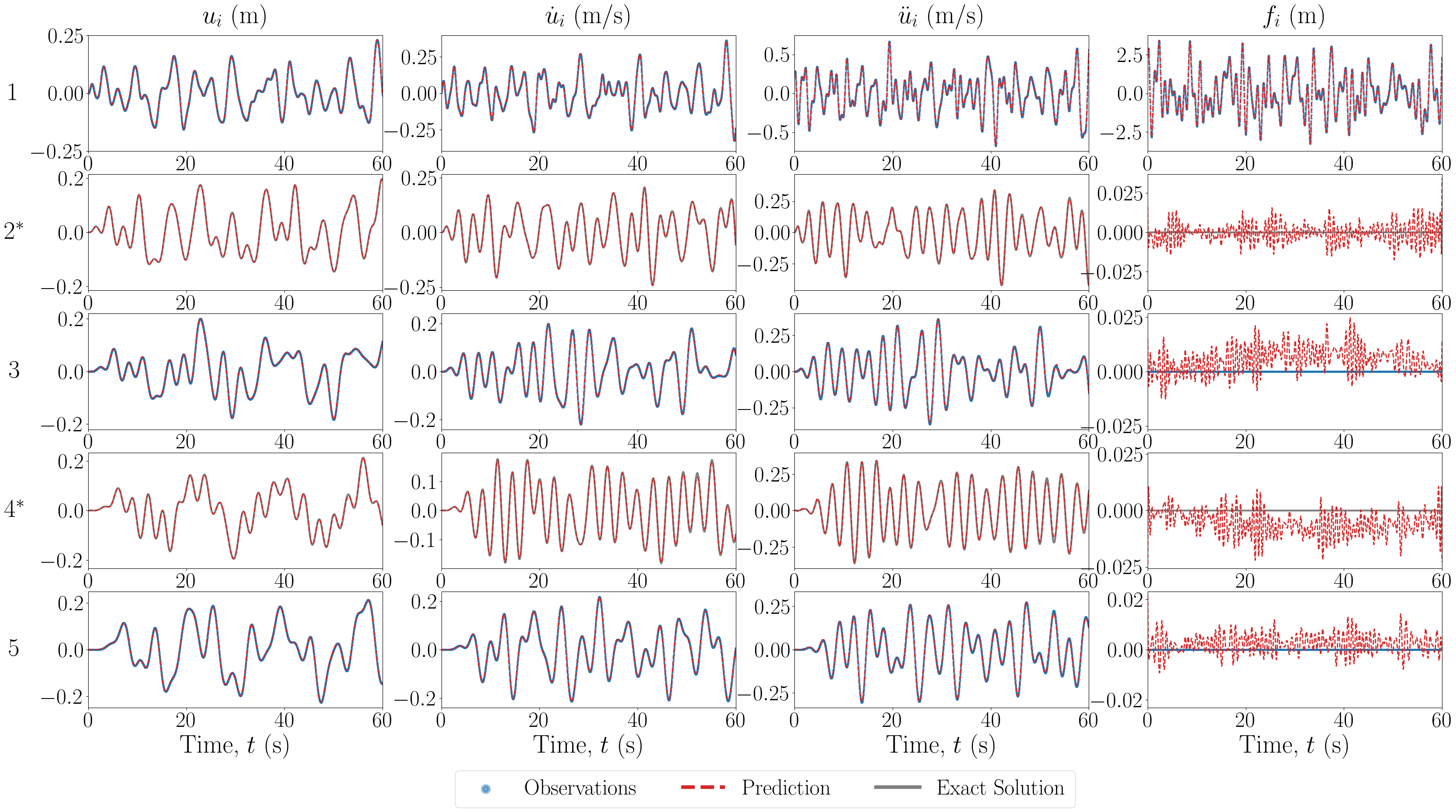}
    \caption{Estimated input and response for 5DOF Van Der Pol system, from the state-parameter estimation scheme, with no model error. The DOF is indicated on the left, with an asterisk indicating unobserved DOFs.}
    \label{fig:spe-nonerr-vdp}
\end{figure}

To confirm the robustness of the state-parameter estimation scheme, the same case was applied to a 5DOF Van Der Pol system. %
The results of the state and force estimation are shown in \Cref{fig:spe-nonerr-vdp}, and the estimated values of the system parameters are shown in \Cref{tab:spe-vdp-err}. %
As above, the estimated response and force signals match well with the ground truth and the estimated values of the system parameters are close to the true values. %
The largest error is 6.886\% for $\mu_2$, which may be explained similarly to the points given above. %

\begin{table}[h!]
    \caption{Values of estimated parameters from state-parameter estimation scheme applied to various MDOF systems with different errors.}\label{tab:spe-vdp-err}
    \begin{tabular*}{\textwidth}{@{\extracolsep\fill}rcccrcccrccc}
        \toprule
        \multicolumn{4}{@{}c@{}}{Van Der Pol} & \multicolumn{4}{@{}c@{}}{Linear Model\footnotemark[1]} & \multicolumn{4}{@{}c@{}}{Value Error} \\
        \cmidrule{1-4}\cmidrule{5-8}\cmidrule{9-12}
         & True & Estimated & Error &  & True & Estimated & Error &  & True & Estimated & Error \\
        \midrule
       $k_2$ & 15.0 & 14.76 & 1.61\% & $k_2$ & 15.0 & 15.70 & 4.60\% & $k_2$ & 13.95 & 13.40 & 3.98\% \\
       $k_4$ & 15.0 & 14.72 & 1.87\% & $k_4$ & 15.0 & 14.95 & 0.34\% & $k_4$ & 13.62 & 13.90 & 2.06\% \\
       $\mu_2$ & 0.75 & 0.802 & 6.89\% &  & & &                       & $\kappa_2$ & 93.02 & 119.25 & 28.2\% \\
       $\mu_4$ & 0.75 & 0.775 & 3.33\% &  & & &                     & $\kappa_4$ & 90.78 & 109.20 & 20.27\% \\
       \botrule
    \end{tabular*}
\footnotetext[1]{The prescribed linear model case study is applied to data from the Duffing oscillator system.}
\end{table}

\begin{figure}[h!]
    \centering
    \includegraphics[width=\linewidth]{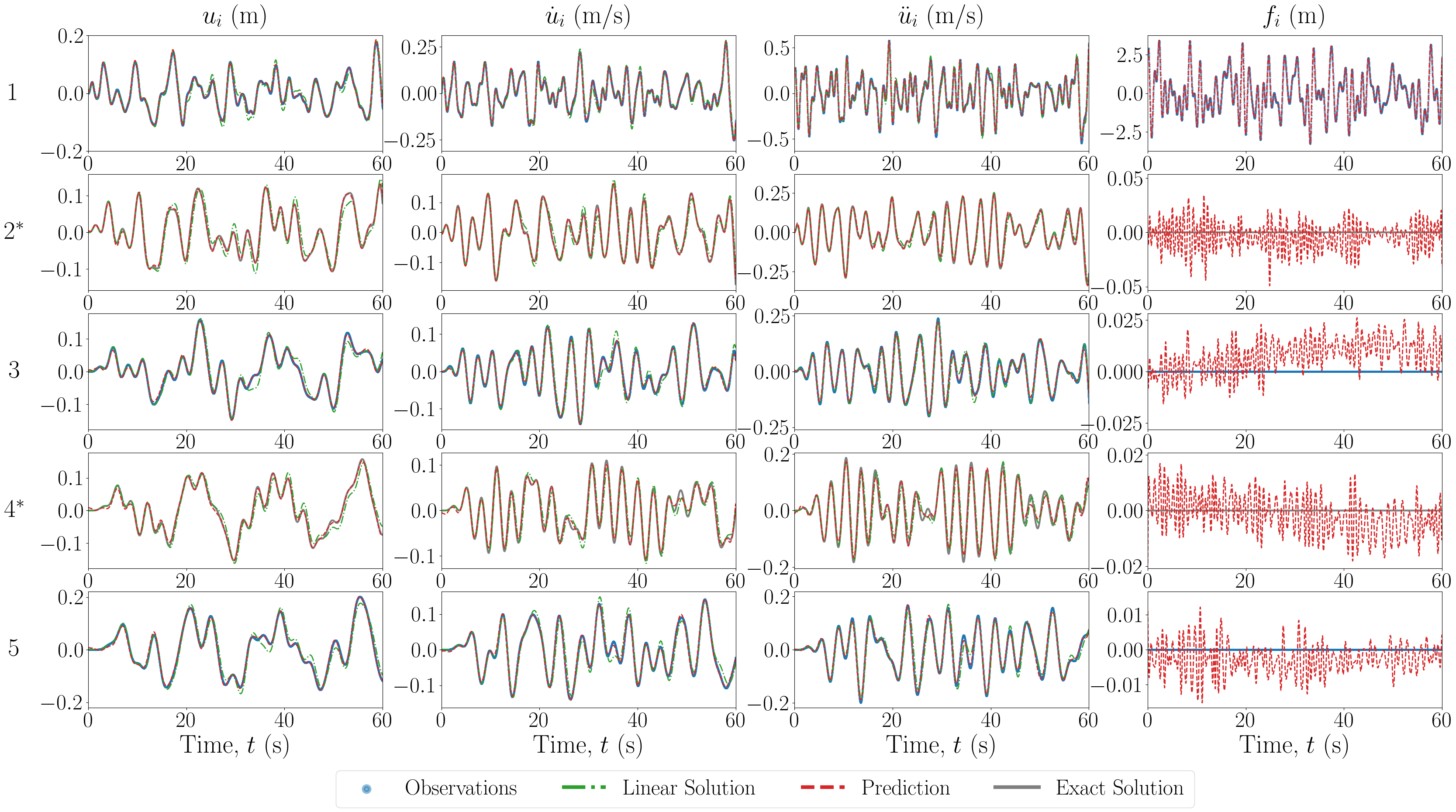}
    \caption{Estimated input and response for 5DOF Duffing system, from the state-parameter estimation scheme, with model error in the form of a prescribed linear model. The DOF is indicated on the left, with an asterisk indicating unobserved DOFs.}
    \label{fig:spe-linmod-duf}
\end{figure}

In the next case scenario, the model prescribed in the PINN is set to be the equivalent linear system, i.e.\ the nonlinear forces are set to zero. %
The results of the state estimation for a 5DOF Duffing system are shown in \Cref{fig:spe-linmod-duf}, along with the true response to the equivalent linear system with the same input. %
Even with the equivalent linear model prescribed, the response estimation is relatively close to the ground truth, especially when compared to the equivalent linear response. %
This could be explained by the fact that the unobserved DOFs are only a maximum of one connection away from the observed DOFs, therefore, the discrepancy in the necessary restoring forces would not propagate. %
The estimated values of the system parameters are shown in \Cref{tab:spe-vdp-err}. %
The accuracy of the estimation of the linear spring is reduced, likely conceptually explained as the estimation of a `best-equivalent' linear constant to match the restoring forces produced by the nonlinear spring. %

\begin{figure}[h!]
    \centering
    \includegraphics[width=\linewidth]{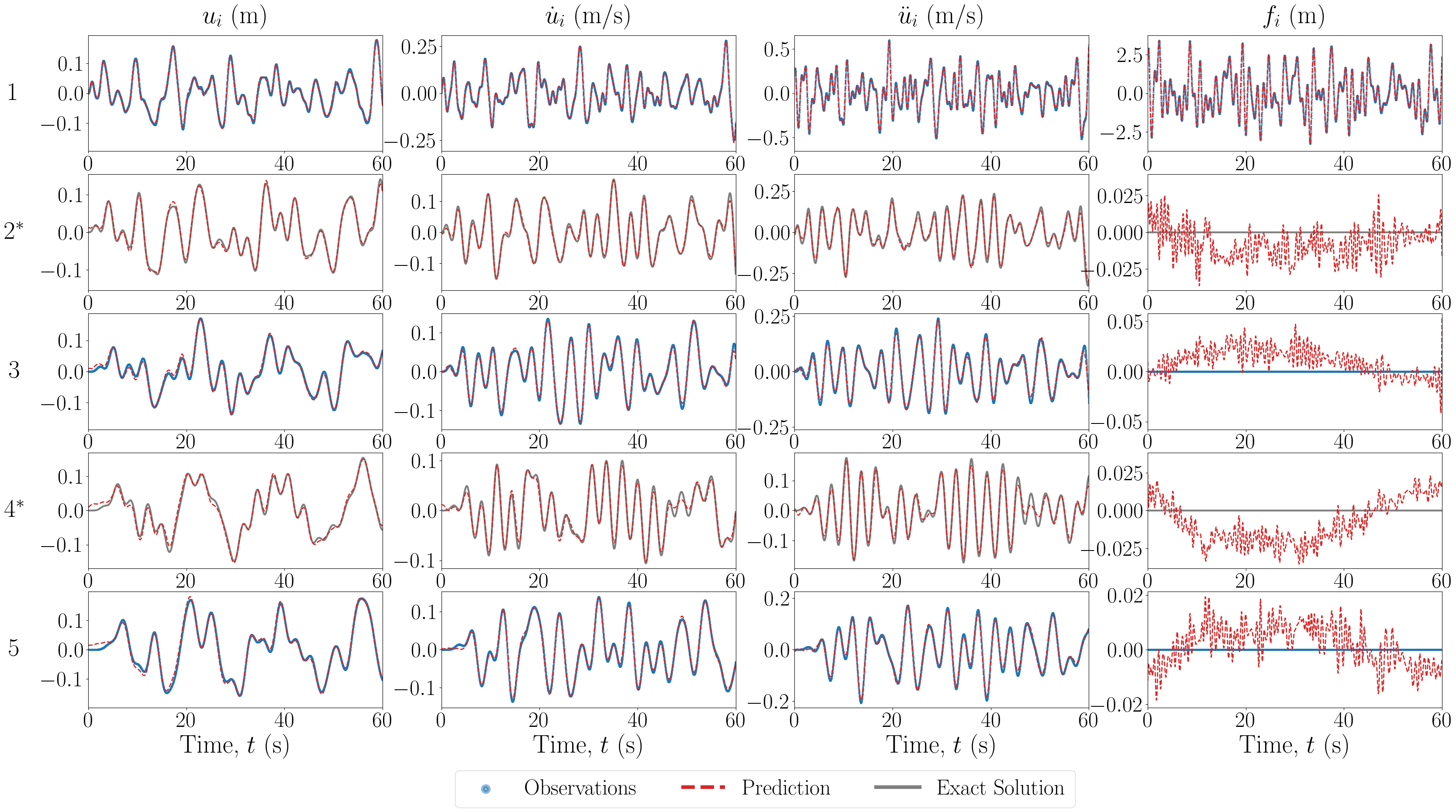}
    \caption{Estimated input and response for 5DOF Duffing system, from the state-parameter estimation scheme, with model error in the form of incorrect system parameter values. The DOF is indicated on the left, with an asterisk indicating unobserved DOFs.}
    \label{fig:spe-valerr-duf}
\end{figure}

In the final case scenario, the model error is of the form of system parameter value discrepancies. %
The results of the state estimation for this case applied to a 5DOF Duffing oscillator are shown in \Cref{fig:spe-valerr-duf}. %
The values of the true and estimated system parameters are shown in \Cref{tab:spe-vdp-err}. %
In this case, the discrepancy in system values does not result in a pronounced error in the response estimation, likely again explained by the fact that the unobserved DOFs are closely spaced to (only one connection away from) observed DOFs. %
However, the error in the estimated parameters has increased when compared to the case where there is no model error. %
This is once again likely explained by the parameters optimisation forcing them to determine a `best-equivalent' set of parameters to match the discrepancy in the restoring forces from the prescribed and true model. %

\Cref{tab:spe-summary} summarises the fitness metrics of the case studies performed for the state-parameter estimation scheme, which are calculated as shown in \Cref{eq:fitness}. %
The fitness of the predictions is, again, generally high, with the lowest fitness in the unobserved domain being 97.66\% for the displacement of the case with model error in the form of a linear model. %
When comparing the fitness of the observed and unobserved degrees of freedom, the unobserved degrees of freedom generally have a lower fitness, but only when there are model errors. %
Furthermore, the fitness is consistently worse for the estimation of the displacement, and is more pronounced when there are modelling errors. %
Similarly to the results of the sparse recovery scheme, the estimation of the forces in the state-parameter estimation scheme match well with the ground truth, with the value of the recovered force at DOFs where the truth is zero being of much smaller magnitude than the true input signal. %

\begingroup
\renewcommand{\arraystretch}{1.2}
\begin{table}[h!]
    \caption{Summary of fitness metrics in percent (\%) of the case studies performed for the state-parameter estimation scheme.}\label{tab:spe-summary}
    \centering
    \begin{tabular*}{\textwidth}{@{\extracolsep\fill}ccccccccccccc}
    \toprule
             & \multicolumn{4}{@{}c@{}}{All ($\Omega_p$)} & \multicolumn{4}{@{}c@{}}{Observed ($\Omega_o$)} & \multicolumn{4}{@{}c@{}}{Unobserved ($\Omega_o$)} \\\cmidrule{2-5}\cmidrule{6-9}\cmidrule{10-13}
        Case & $\mathbf{u}$ & $\dot{\mathbf{u}}$ & $\ddot{\mathbf{u}}$ & $\mathbf{f}$ & $\mathbf{u}$ & $\dot{\mathbf{u}}$ & $\ddot{\mathbf{u}}$ & $\mathbf{f}$ & $\mathbf{u}$ & $\dot{\mathbf{u}}$ & $\ddot{\mathbf{u}}$ & $\mathbf{f}$ \\ 
        \midrule
        \begin{tabular}[c]{c}5DOF - Duffing\\No Error\\SNR: 50.0\end{tabular} & 99.60 & 99.72 & 99.64 & - & 99.65 & 99.80 & 99.72 & 99.73 & 99.53 & 99.59 & 99.52 & -    \\
        \hline
        \begin{tabular}[c]{c}5DOF - Duffing\\No Error\\SNR: 20.0\end{tabular} & 98.85 & 99.24 & 99.38 & - & 98.97 & 99.34 & 99.56 & 99.41 & 98.66 & 99.07 & 99.11 & -     \\
        \hline
        \begin{tabular}[c]{c}5DOF - Duffing\\No Error\\SNR: 10.0\end{tabular} & 97.15 & 98.12 & 98.40 & - & 97.37 & 98.37 & 98.79 & 98.361 & 96.81 & 97.73 & 97.82 & -     \\
        \hline
        \begin{tabular}[c]{c}5DOF - Van Der Pol\\No Error\\SNR: 50.0\end{tabular} & 98.63 & 99.03 & 99.17 & - & 98.77 & 99.12 & 99.34 & 98.42 & 98.41 & 98.90 & 98.91 & -     \\
        \hline
        \begin{tabular}[c]{c}5DOF\\Linear Model\\SNR: 50.0\end{tabular} & 98.00 & 98.77 & 99.10 & - & 98.23 & 99.18 & 99.48 & 99.31 & 97.66 & 98.16 & 98.54 & - \\
        \hline
        \begin{tabular}[c]{c}5DOF\\Value Error\\SNR: 50.0\end{tabular} & 98.56 & 99.01 & 99.36 & - & 98.81 & 99.45 & 99.79 & 99.71 & 98.18 & 98.35 & 98.72 & - \\
    \botrule
    \end{tabular*}
\footnotetext{Note: No fitness score is available for some force vector estimates as the true signal is zero in these locations (unforced).}
\end{table}

\subsection{Parameter Estimation}
For the parameter estimation scheme, a time window of 240s with 4096 samples was simulated, and noise was added to the observation signals corresponding to a SNR of 20, and repeated 5 times. %
These 5 noisy signals were then used as observations for the neural network. %
Furthermore, for this case study, values of the state $\mathbf{z}^*$ were used directly as observations in the training of the NN, as no prior knowledge of \emph{any} system parameters would create an under-defined problem if only given observations of acceleration $\ddot{\mathbf{u}}^*$. %

The results of 5000 iterations of the HMC sampling procedure applied to a 2DOF Duffing system are shown in \Cref{fig:mcmc-results-duf}. %
The diagonal elements of the gridded array show the univariate sample distributions for the $i^{\textrm{th}}$ parameter, the lower triangle show the multivariate sample distributions comparing the $i^{\textrm{th}}$ and $j^{\textrm{th}}$ parameter, and the upper triangle gives the value of correlation between the $i^{\textrm{th}}$ and $j^{\textrm{th}}$ parameter. %
\Cref{tab:mcmc-results} shows the arithmetic mean and standard deviation of the parameters for the Duffing oscillator sampled using the HMC procedure, as well as the relative error of the mean compared to the true value. %

From \Cref{fig:mcmc-results-duf} one can see the strong correlation between the linear and nonlinear stiffness at the same connection points (i.e. $k_1$ with $\kappa_1$ and $k_2$ with $\kappa_2$), which is to be expected as they are both dependent on the relative displacement of that connection. %
Other than this, there appears no correlation between parameters. %
This matches well with prior knowledge on the correlated distribution of parameters for a nonlinear MDOF oscillator \cite{rogers_latent_2022,brynjarsdottir_learning_2014}. %

\begin{figure}[h!]
    \centering
    \includegraphics[width=\linewidth]{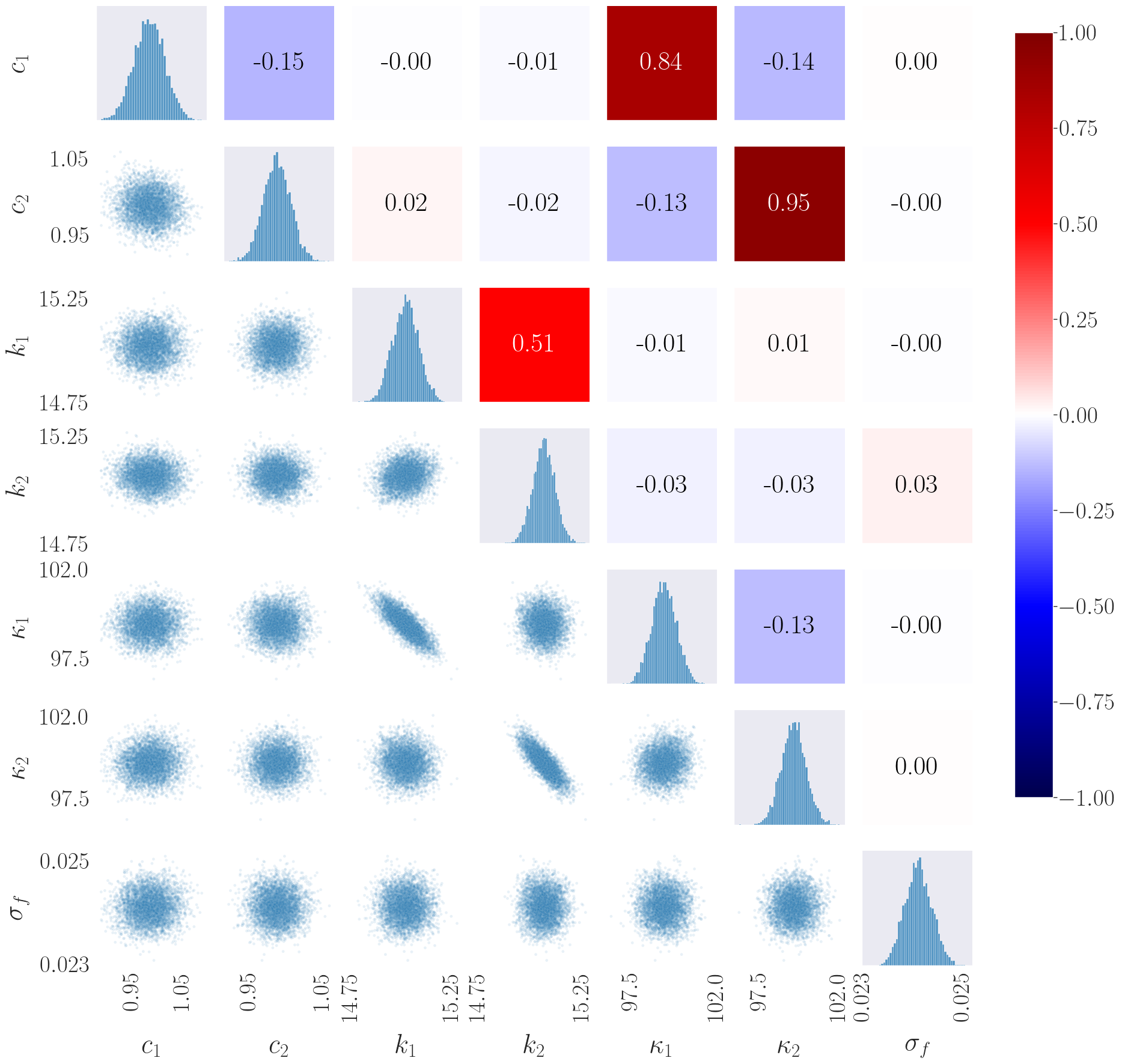}
    \caption{Posterior samples of MCMC procedure applied over uncertain parameters of 2DOF Duffing oscillator, and the sample correlation between parameters.}
    \label{fig:mcmc-results-duf}
\end{figure}

\begin{table}[h!]
    \caption{Value and relative error of the calculated mean and standard deviation of posterior samples of unknown parameters of 2DOF Duffing oscillator}
    \label{tab:mcmc-results}
    \begin{tabular*}{\textwidth}{@{\extracolsep\fill}lcccccccc}
        \toprule
        & \multicolumn{4}{@{}c@{}}{Duffing} & \multicolumn{4}{@{}c@{}}{Van Der Pol} \\
        \cmidrule{2-5}\cmidrule{6-9}
        Parameter & True Value & Mean & Error & STD\footnotemark[1] & True Value & Mean & Error & STD\footnotemark[1] \\
        \midrule
        $c_1$ & 1.0 & 0.984       & 1.60\%    & 0.030 & 1.0 & 1.05  & 5.00\%    & 0.107 \\
        $c_2$ & 1.0 & 0.987       & 1.30\%     & 0.030 & 1.0 & 0.998 & 0.20\%    & 0.132 \\
        $k_1$ & 15.0 & 15.02       & 0.13\%    & 0.065 & 15.0 & 14.92 & 0.53\%    & 0.059 \\
        $k_2$ & 15.0 & 15.06       & 0.43\%    & 0.053 & 15.0 & 14.85 & 1.03\%    & 0.062 \\
        $\kappa_1$/$\mu_1$ & 100.0 & 99.21  & 0.79\%    & 0.65 & 0.75 & 0.720 & 5.00\%    & 0.132 \\
        $\kappa_2$/$\mu_2$ & 100.0 & 99.43  & 0.57\%   & 0.70 & 0.75 & 0.788 & 5.07\%    & 0.156 \\
        \botrule
    \end{tabular*}
    \footnotetext[1]{STD is the calculated arithmetic standard deviation.}
\end{table}


For all stiffness parameters (linear and nonlinear), and linear damping parameters, the estimated means appear to fit well to the true value, and the standard deviations are not large. %
However, the standard deviation for the nonlinear damping parameters are slightly worse, and the relative error of the mean value is also quite high. %
As the relative amplitude of the restoring force from the nonlinear damper is more sensitive to the states, the model may be overfitting to the data, and thus the uncertainty in the parameter estimation is increased. %

\Cref{fig:mcmc-results-vdp} shows the posterior samples of the HMC procedure applied to a 2DOF Van Der Pol oscillator, where the layout is the same as in \Cref{fig:mcmc-results-duf}. %
Similarly to the linear and nonlinear stiffness components of the Duffing system, there appears to be a strong correlation between the linear and nonlinear `damping' components at the same connection points (i.e.\ $c_1$ with $\mu_1$ and $c_2$ with $\mu_2$). %
However, in these results, the correlation is positive, instead of the negative correlation seen in \Cref{fig:mcmc-results-duf}. %
This may be a result of the dependence of the Van Der Pol damping on both the velocity and the displacement, and so phase will play a role in the interdependencies between these parameters. %

\begin{figure}[h!]
    \centering
    \includegraphics[width=\linewidth]{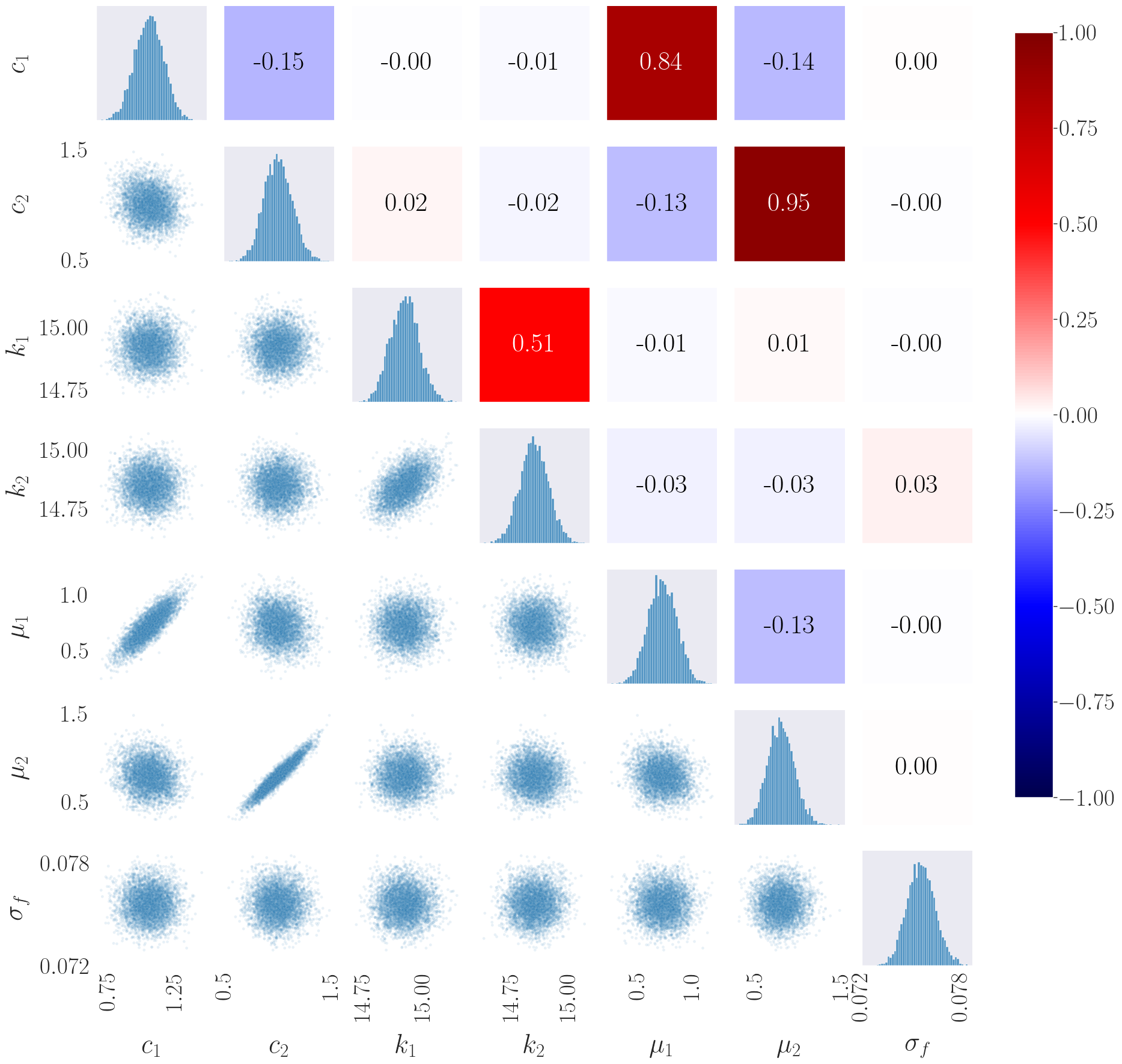}
    \caption{Posterior samples of MCMC procedure applied over unknown parameters of 2DOF Van Der Pol oscillator, and the sample correlation between parameters.}
    \label{fig:mcmc-results-vdp}
\end{figure}

The calculated mean and standard deviation for the 2DOF Van Der Pol oscillator, along with the relative error of the mean, are given in \Cref{tab:mcmc-results}. %
From the values of the relative error, the HMC sampler appears to have not performed as well at estimating the mean value for the Van Der Pol oscillator when compared to the Duffing system. %
This may be a result of the dependency of the nonlinearity on both the displacement and velocity, and so more information is required for an accurate result, such as a longer time window. %
Furthermore, the standard deviations of the damping parameters $c_i$ are larger, indicating an increased uncertainty in the value. %
This larger uncertainty is not translated to the linear stiffness parameter. %

\section{Discussion}


The results of the investigated case studies demonstrate that the PINN can be applied for multiple downstream tasks that are often motivated with dynamic structures. %
When performing solely the sparse recovery scheme, the PINN is able to accurately predict the response of the system, even when there are model errors. %
The authors here emphasise again, however, that the PINN can only function in batch mode, and would, thus, not be suited for real-time (online) estimation tasks, where step ahead prediction is of the essence. %
Different modelling errors, i.e.\ whether they are due to incorrect model form or inaccurate system parameter values, do not seem to incur significantly different effects on the accuracy of the prediction of the state. %
As the setup generates a prediction at each time point independently of others, the PINN is still able to accurately predict the response of the system, even under the presence of modelling errors, as there is no propagation of error. %
When performing sparse recovery in a domain extension context, the prediction of the response begins to worsen as the size of the attempted extension is increased, even when the extended region is still the same relative size to the observed domain. %
This is likely because of the fact that the input is only applied to the first degree of freedom, and so there is another limitation in the current setup with assuming zero-forces elsewhere, presenting another avenue of research for this problem. %

When performing the state-parameter estimation scheme, the PINN is able to accurately simultaneously predict the response and parameters of the system. %
The model errors do have a more pronounced effect on the estimation of the parameters, as the optimisation is forced to determine a `best-equivalent' set of parameters to match the discrepancy in the restoring forces from the prescribed and true model. %
As the state-parameter estimation setup involved having an interpolated recovery scheme, the state estimation was generally more accurate than the domain extension approach, which is an intuitive results, but worthy to note. %

In the parameter estimation scheme, it was shown how PINN architectures could be exploited for probabilistic representations of parameter estimation using sampling schemes such as HMC. %
Generally, the approach performed well to generate an appropriate set of posterior samples of the system parameters, and the captured correlations between parameters is as expected from other literature. %
The advantage of such a scheme applied in this orientation, is that for system with complex physics (such as Van Der Pol oscillators), generating a proposed solution to the equation for such sampling schemes can be computationally intense, reducing the applicability of such an approach. %
Whereas the PINN, once trained on the observation data, can rapidly reproduce an estimate of the observed forcing signal, given a new set of parameters. %


Generally, modelling errors did not have a significant effect on the prediction of the response, but did have a more pronounced effect on the estimation of the parameters. %
Between having inaccurate parameters vs having an inaccurate model form, the former had a more pronounced effect on the estimation of the parameters. %
Therefore, when utilising PINNs for either of the schemes, as the objective is in a weak form, missing terms in the physics model may not result in large discrepancies in prediction, however, this is likely dependent on the relative effect of the missing term. %


When applying the PINN for dynamic systems, the main challenge lies in the determination of the hyperparameters and network architecture design. %
There is little intuition on the appropriate number of layers and nodes, and the choice of activation functions, which can significantly affect the performance of the model. %
Furthermore, the weighting of the losses is extremely important, and the success of model conversion is highly dependent on these values. %

The limitations of the approach given here is that the model is very non-generalised, and is highly dependent on the specific system and instance being modelled. %
For example, if one were to wish to predict the next state of a system given its current state, the model would have to be retrained for each new instance, which is obviously impractical. %
However, for such an example, one could use a setup involving a recurrent neural network, which is more suited to such a task \cite{haywood2023physics}. %


In future work, we plan to investigate the effect of the relative nonlinearity of the system on the performance of the PINN. %
As described above, the prediction capability even with model errors may only be due to the relative linearity of the system. %
It would also be useful to investigate further methods of generalising the model, such that rapid predictions can be made for new instances of the system, without the need for retraining. %
The authors also intend to pursue extending the probabilistic representation of the model, such that the uncertainty in the model can be better represented, even when performing more than just parameter estimation. %
Quantification of uncertainty often requires multiple samples of the same instance, which can be impractical for structural assessment \emph{in-situ}, particularly for structures which often use ambient signals such as bridges. %
Inclusion of physics as a knowledge prior can benefit by allowing a reasonable estimation of noisy processes with only a single measurement. %

\section{Conclusion}

This paper presents the application of a physics-informed neural network (PINN) for the sparse recovery, state-parameter estimation, and parameter estimation of multi-degree-of-freedom (MDOF) systems. %
The PINN is able to accurately predict the response of the system in an offline setting, even in the presence of modelling errors, inaccurate parameters, and measurement noise, and is able to simultaneously predict the response and parameters of the system. %
The model errors do have a more pronounced effect on the estimation of the parameters, as the optimisation is forced to determine a `best-equivalent' set of parameters to match the discrepancy in the restoring forces from the prescribed and true model. %
The results of the case studies show that the PINN can be applied for multiple downstream tasks that are often motivated with dynamic structures. %
Future work has been discussed on improving generalisation, and the authors intend to pursue extending the probabilistic representation of the model. %

\backmatter

\bmhead{Supplementary information}

All the code and data for this article is available open access at a Github repository available at \hyperlink{https://github.com/MarcusHA94/structural-dynamics-pinns}{https://github.com/MarcusHA94/structural-dynamics-pinns}. %

\bmhead{Acknowledgements}

The authors gratefully acknowledge the funding from the Swiss National Science Foundation (SNSF) under the Horizon Europe funding guarantee, for the project ‘ReCharged - Climate-aware Resilience for Sustainable Critical and interdependent Infrastructure Systems enhanced by emerging Digital Technologies’ (grant agreement No: 101086413), and funding from the French National Research Agency (ANR PRCI Grant No. 266157) and the Swiss National Science Foundation (Grant No. 200021L\_212718) for the MISTERY project.

\section*{Declarations}

The authors declare no conflicts of interest.

\begin{appendices}

\section{System Matrices}\label{secA1}

The matrices for the generic MDOF Duffing system used in this paper are given by,
\begin{align*}
    \mathbf{M} &= 
    \begin{bmatrix}
        m_1 & 0 & 0 & ... & 0 \\
        0 & m_2 & 0 & ... & 0 \\
        0 & 0 & m_3 & ... & 0 \\
        \vdots & \vdots & \vdots & \ddots & \vdots \\
        0 & 0 & ... & 0 & m_n
    \end{bmatrix},
    \qquad
    \mathbf{C} = 
    \begin{bmatrix}
        c_1 + c_2 & -c_2 & 0 & ... & 0 \\
        -c_2 & c_2 + c_3 & -c_3 & ... & 0 \\
        0 & -c_3 & c_3 + c_4 & ... & 0 \\
        \vdots & \vdots & \vdots & \ddots & \vdots \\
        0 & 0 & ... & -c_{n-1} & c_n
    \end{bmatrix} \\
    \mathbf{K} &= 
    \begin{bmatrix}
        k_1 + k_2 & -k_2 & 0 & ... & 0 \\
        -k_2 & k_2 + k_3 & -k_3 & ... & 0 \\
        0 & -k_3 & k_3 + k_4 & ... & 0 \\
        \vdots & \vdots & \vdots & \ddots & \vdots \\
        0 & 0 & ... & -k_{n-1} & k_n
    \end{bmatrix},
    \qquad
    \mathbf{C}_n = 
    \begin{bmatrix}
        \mu_1 & -\mu_2 & 0 & ... & 0 \\
        0 & \mu_2 & -\mu_3 & ... & 0 \\
        0 & 0 & \mu_3 & ... & 0 \\
        \vdots & \vdots & \vdots & \ddots & \vdots \\
        0 & 0 & ... & 0 & \mu_n
    \end{bmatrix} \\
    \mathbf{K}_n &= 
    \begin{bmatrix}
        \kappa_1 & -\kappa_2 & 0 & ... & 0 \\
        0 & \kappa_2 & -\kappa_3 & ... & 0 \\
        0 & 0 & \kappa_3 & ... & 0 \\
        \vdots & \vdots & \vdots & \ddots & \vdots \\
        0 & 0 & ... & 0 & \kappa_n
    \end{bmatrix}
\end{align*}

\end{appendices}


\bibliography{mdof-pinn-bib}

\end{document}